\documentclass[sigconf]{acmart}

\usepackage{booktabs} 
\usepackage{soul}
\usepackage{hyperref}
\usepackage{balance}
\newcommand{\eg}{e.\,g.\,, }
\newcommand{\ie}{i.\,e.\,, }

\newcommand{\cf}{{cf.\,}}

\setcopyright{rightsretained}

\acmDOI{10.475/123_4}

\acmISBN{123-4567-24-567/08/06}

\acmConference[ACM MM'19]{ACM Multimedia conference}{October 2019}{Nice, France}
\acmYear{2019}
\copyrightyear{2019}
\acmPrice{15.00}

\begin{document}
\title{AVEC 2019 Workshop and Challenge: State-of-Mind, Detecting Depression with AI, and Cross-Cultural Affect Recognition}

\setcopyright{acmcopyright}
\acmConference[AVEC'19]{2019 Audio/Visual Emotion Challenge and
Workshop}{October, 2019}{Nice, France}
\acmBooktitle{2019 Audio/Visual Emotion Challenge and Workshop
(AVEC'19), October, 2019, Nice, France}
\acmDOI{10.1145/3266302.3266316}
\acmISBN{978-1-4503-5983-2/18/10}

\author{Fabien Ringeval}
\orcid{0000-0002-9213-4529}
\affiliation{
  \institution{Universit\'e Grenoble Alpes, CNRS}
  \city{Grenoble}
  \country{France}
}

\author{Bj\"orn Schuller}
\authornote{The author is further affiliated with Imperial College London, London, UK.}
\affiliation{
  \institution{University of Augsburg}
  \city{Augsburg}
  \country{Germany}}

\author{Michel Valstar}
\affiliation{
  \institution{University of Nottingham}
  \city{Nottingham}
  \country{UK}}

\author{Nicholas Cummins}
\affiliation{
  \institution{University of Augsburg}
  \city{Augsburg}
  \country{Germany}}

\author{Roddy Cowie}
\affiliation{
  \institution{Queen's University Belfast}
  \city{Belfast}
  \country{UK}}

\author{Leili Tavabi}
\affiliation{
  \institution{University of Southern California}
  \city{Los Angeles}
  \country{USA}}

\author{Maximilian Schmitt}
\affiliation{
  \institution{University of Augsburg}
  \city{Augsburg}
  \country{Germany}}

\author{Sina Alisamir}
\affiliation{
  \institution{Universit\'e Grenoble Alpes, CNRS}
  \city{Grenoble}
  \country{France}}

\author{Shahin Amiriparian}
\affiliation{
  \institution{University of Augsburg}
  \city{Augsburg}
  \country{Germany}}

\author{Eva-Maria Messner}
\affiliation{
  \institution{University of Ulm}
  \city{Ulm}
  \country{Germany}}

\author{Siyang Song}
\affiliation{
  \institution{University of Nottingham}
  \city{Nottingham}
  \country{UK}}

\author{Shuo Liu}
\affiliation{
  \institution{University of Augsburg}
  \city{Augsburg}
  \country{Germany}}

\author{Ziping Zhao}
\affiliation{
  \institution{Tianjin Normal University}
  \city{Tianjin}
  \country{China}}

\author{Adria Mallol-Ragolta}
\affiliation{
  \institution{University of Augsburg}
  \city{Augsburg}
  \country{Germany}}

\author{Zhao Ren}
\affiliation{
  \institution{University of Augsburg}
  \city{Augsburg}
  \country{Germany}}

\author{Mohammad Soleymani}
\affiliation{
  \institution{University of Southern California}
  \city{Los Angeles}
  \country{USA}}
  
\author{Maja Pantic}
\authornote{The author is further affiliated with University of Twente, Twente, The Netherlands.}
\affiliation{%
 \institution{Imperial College London}
 \city{London}
 \country{UK}}

\renewcommand{\shortauthors}{F. Ringeval et al.}
\renewcommand{\shorttitle}{AVEC 2019: State-of-Mind, Detecting Depression with AI, and Cross-Cultural Affect}

\begin{abstract}
The Audio/Visual Emotion Challenge and Workshop (AVEC 2019) ``State-of-Mind, Detecting Depression with AI, and Cross-cultural Affect Recognition'' is the ninth competition event aimed at the comparison of multimedia processing and machine learning methods for automatic audiovisual health and emotion analysis, with all participants competing strictly under the same conditions. The goal of the Challenge is to provide a common benchmark test set for multimodal information processing and to bring together the health and emotion recognition communities, as well as the audiovisual processing communities, to compare the relative merits of various approaches to health and emotion recognition from real-life data. This paper presents the major novelties introduced this year, the challenge guidelines, the data used, and the performance of the baseline systems on the three proposed tasks: state-of-mind recognition, depression assessment with AI, and cross-cultural affect sensing, respectively.
\end{abstract}

%
%



\maketitle

\section{Introduction}
\label{intro}
\noindent 
The Audio/Visual Emotion Challenge and Workshop (AVEC 2019) is the ninth competition aimed at comparison of multimedia processing and machine learning methods for automatic audio, visual, and audiovisual health and emotion sensing, with all participants competing strictly under the same conditions~\cite{Schuller11-A2T, Schuller12-A2T, Valstar13-A2T, Valstar14-A2T, Ringeval15-A2Ta, Valstar16-SFA, Ringeval17-SFA, Ringeval18-SFA}. 

One of the goals of the AVEC series is to bring together multiple communities from different disciplines, in particular, the audiovisual multimedia communities and those in the psychological and social sciences who study expressive behaviour. 
Another objective is to advance health and emotion recognition systems by providing a common benchmark test set for multimodal information processing, in order to compare the relative merits of the approaches to automatic health and emotion analysis under well-defined conditions, \ie 
with large volumes of un-segmented, non-prototypical and non-preselected data of wholly naturalistic behaviour. This is precisely the type of data that the new generation of affect-oriented multimedia and human-machine/human-robot communication interfaces have to face in the real world.

Major novelties are introduced for the AVEC 2019 with three separated Sub-challenges focusing on health and emotion analysis: (i) State-of-Mind Sub-challenge (SoMS), (ii) Detecting Depression with AI Sub-challenge (DDS), and (iii) Cross-cultural Emotion Sub-challenge (CES). In the following, we describe the novelties introduced in the Challenge and the guidelines for participating. 

The State-of-Mind Sub-challenge (SoMS) is a new task focusing on the continuous adaptation of human state-of-mind (SOM), which is pivotal for mental functioning and behaviour regulation~\cite{Houben15-TRB}. 
SOM is constantly shifting due to internal and external stimuli, and 
frequent use of either adaptive or maladaptive SOM influences our mental health. One key aspect of the human experience is the way emotion features in our SOM~\cite{shapiro02,Schwarz83-MMA}. In the SoMS, self-reported mood (10-point Likert scale) after the narrative of personal stories (two positive and two negative), has to be predicted automatically from the audiovisual recordings of those stories; USoM corpus~\cite{rathner18som}. 

The Detecting Depression with AI Sub-challenge (DDS) is a major extension of the AVEC 2016 DSC~\cite{Valstar16-A2D}, where the level of depression severity (PHQ-8 questionnaire) was assessed from audiovisual recordings of patients interacting with a virtual agent conducting a clinical interview and driven by a human as a Wizard-of-Oz (WoZ); DAIC-WOZ corpus~\cite{gratchetal14}. The DAIC data set contains new recordings of the same population with the virtual agent being, this time, wholly driven by AI, \ie without any human intervention. Those new recordings are used as a test partition for the DDS, and will help to understand how the absence of a human conducting the virtual agent impacts on automatic depression severity assessment.

The Cross-cultural Emotion Sub-challenge (CES) is a large extension of the AVEC 2018 CES~\cite{Ringeval18-A2W}, where dimensions of emotion were inferred from audiovisual recordings collected ``\emph{in-the-wild}'', \ie with standard webcams and at home/work place. A cross-cultural setup was further exploited for inferring emotion: knowledge of German culture was leveraged to infer emotion on the Hungarian culture, using the SEWA corpus~\cite{kossaifi19-sda}. This dataset now includes data collected from new participants with Chinese culture, which is used as a test set for the CES, whose aim is, therefore, to investigate how emotion knowledge of Western European cultures (German, Hungarian) can be transferred to the Chinese culture. 

All Sub-challenges allow contributors to find their own features to use with their own machine learning algorithm. In addition, standard feature sets are provided for audio and video data (\cf Section~\ref{Baseline_features}), along with scripts available in a public repository\footnote{\url{https://github.com/AudioVisualEmotionChallenge/AVEC2019}}, which participants are free to use for reproducing both the baseline features and recognition systems (\cf Section~\ref{Baseline_systems}). The labels of the test partition remain unknown to the participants, and participants have to stick to the definition of training, development, and test partition. They may freely report on results obtained on the development partition, but are limited to five trials per Sub-challenge in submitting their results on the test partition. 

Ranking of the labels 
relies on the \emph{Concordance Correlation Coefficient ($CCC$)}~\cite{Li89-ACC} for all Sub-challenges; the Root Mean Squared Error ($RMSE$) is additionally reported. Whereas many other metrics of performance could be exploited for ranking the contributions, such as the Spearman's $CC$, or the coefficient of determination ($r^2$), we believe that the index of reproducibility \emph{$CCC$} is the most suitable metric to use, as it is not biased by changes in scale and location, and elegantly includes information on both precision and accuracy in a single statistical measure~\cite{Li89-ACC}. Moreover, its theoretical definition and properties are well rooted in the literature~\cite{Pandit19-OMM}, and it can be easily exploited as a loss function for training neural networks~\cite{Weninger16-DTR}.

To be eligible to participate in the Challenge, every entry has to be accompanied by a paper submitted to the AVEC 2019 Data Challenge and Workshop, describing the results and the methods that created them. These papers undergo peer-review by the technical program committee. Only contributions with a relevant accepted paper and at least a submission of test results are eligible for participation. The organisers do not participate in the Challenge themselves, but re-evaluate the findings of the best performing system of each Sub-challenge. 


The remainder of this paper is organised as follows. We summarise relevant related work in Section~\ref{related}, introduce the Challenge corpora in Section~\ref{Challenge_corpora}, the common audiovisual baseline feature sets in Section~\ref{Baseline_features}, and the developed baseline recognition systems with the obtained results in Section~\ref{Baseline_systems}, before concluding in Section~\ref{Conclusion}.

\section{Related Work}
\label{related}
This section is a summary of the current state-of-the-art in the automatic analysis of affect with a focus on: (i) human state-of-mind, (ii) depression assessment in the context of AI-driven virtual agents, and (iii) dimensional analysis in cross-cultural paradigms.

\subsection{State-of-Mind}
The concept of a human SOM describes the phenomenon that our consciousness and emotions are constantly fluctuating over time; this is due to internal and external biological, psychological, and social demands~\cite{Houben15-TRB, rathner18som}. One key aspect of SOM is our emotions. They provide valuable information that influences our basic human processes in a bidirectional manner~\cite{shanetal2009,shapiro02}.
Such processes include attention, perception, cognition, memory retrieval, memory storage, and behaviour regulation. In fact, depending on our actual SOM, some emotions, cognitions, and behaviours are more likely to occur, while others may be suppressed. This effect is the underlying principle of mood congruence~\cite{Schwarz83-MMA, russell2003core}. 

Despite the major impact of SOM on health and social functioning, the quantification of current emotional states, with therapy contexts, has its pitfalls. The simplest of these is that it relies heavily on self-reports of emotional states, which are inherently biased~\cite{Yannakakis17-TON}. As humans are structurally determined closed systems, it is not always sound to assume that people who give the same scores on measurement scales are actually in the same SOM~\cite{maturana1987tree}. 
Moreover, even within a person, the current rating of the emotional state is rooted in previous experiences, known as the adaption level, and therefore is not really accurate in an absolute way~\cite{russell1984adaptation}. 

Approaches like Russell's avoid having to limit the quantification to
a given language~\cite{russell2003core}: on his theory of core affect, every instance of
emotion can be quantified on the orthogonal axes \emph{arousal} (from sleepy to
hyper-aroused) and \emph{valence} (with the poles negative and positive). However,
raters' ability to quantify reliably is still doubtful. One approach to 
overcoming this is to treat emotional state values as ordinal variables~\cite{Yannakakis17-TON}. Another is to complement self-ratings with expert ratings or physiological recordings. Each of these methods has its limitations; the mismatch between different emotion assessments is still very much a matter of scientific discourse~\cite{schwerdtfeger2016ecological, schwerdtfeger2004predicting}. 

Despite the given limitations of the scientific assessment of emotional states, humans constantly monitor their own and others' emotions and organise themselves within social systems~\cite{sapolsky2004social, dautenhahn2002origins}. Given the need for humans to socially interact and the increased occurrence of human-machine-interactions, the development of a real-time SOM data-driven recognition system has the potential to enhance user experience, user satisfaction, and subsequently to foster user adherence~\cite{rathner18som, rathner2018did, baumel2018predicting}. Such a system could assist society in various ways; i) decreasing bias in the monitoring of SOM; ii) collecting more objective data to aid the diagnosis of affective disorders; iii) delivering tailored interventions to facilitate treatment of disease; iv) reducing the time spent in the evaluation of treatment outcome, and in e-treatment by presenting SOM related content, easing burdens on both patient and provider~\cite{rathner2018did, rathner18som, schuller2018interspeech, Stappen19-CMU}.

\subsection{Depression Detection with AI}
Depression, particularly major depressive disorder (MDD), is a common mental health problem, with negative impacts on the way one thinks, feels, and acts~\cite{APA13-DSM}. It can lead to a variety of emotional and physical problems and affect many aspects of both working and personal life. The World Health Organisation (WHO) declared depression as the leading cause of ill health and disability worldwide in 2015: more than 300 million people live with it~\cite{world2017depression}. Given the high prevalence of depression and its suicide risk, finding new methods for diagnosis and treatment becomes more and more critical. 

There is growing interest in using automatic human behaviour analysis for computer-aided depression diagnosis based on behavioural cues such as facial expressions and speech prosody, because of convincing evidences that depression and related mental health disorders are associated with changes in patterns of behaviour~\cite{cohn2009detecting, scherer2014automatic, joshi2013multimodal, cummins2015review, williamsonetal13_vbd}. Facial activity, gesturing, head movements and expressivity are among behavioural signals that are strongly correlated with depression. 

Early paralinguistic investigations into depressed speech found that patients consistently demonstrated prosodic speech abnormalities such as reduced pitch, reduced pitch range, slower speaking rate, and higher articulation errors~\cite{cummins2015review}. Facial expression and head gestures that can be tracked by computer vision are also good predictors of depression; e.\,g., a more downward angle of the gaze, less intense smiles, and shorter average duration of smiles have been reported as the most salient facial cues of depression~\cite{SchererEtAl13_ABD}. 
Further, body expressions, gestures, head movements, and linguistic cues have also been reported to provide relevant cues for depression detection~\cite{morales2017cross,ramirez2008psychology, pennebaker2003psychological, althoff2016large}.

Taking all those evidences together, it has been proposed to integrate affective computing technology into a computer agent that interviews people and identifies verbal and nonverbal indicators of mental illnesses~\cite{DeVaultEtAl2014_SVH}. Data collected with subjects suffering from post-traumatic stress disorder showed that the automatic evaluation of their level of depression severity (PHQ-8 questionnaire) can achieve a $RMSE$ less than 5 when the agent is driven by a human acting as a WoZ~\cite{Gong17-TMB}; PHQ-8's range $\in [0, 24]$ and cutpoints are defined at $[5,10,15,20]$ for mild, moderate, moderately severe, and severe depression, respectively. Those results need to be investigated further, with the agent being wholly driven by AI, as the wizard might drive the virtual agent to a situation that eases the observation of patterns associated with depression, or the autonomous agent might have issues in conducting the interview appropriately. 

\subsection{Cross-cultural Emotion Recognition}
Cross-cultural emotion recognition has long been highlighted as an open research question within the affective computing community~\cite{Dmello15-ARA, Elfenbein02-otu, Esposito15-NAC, Pantic05-AMH}, and was introduced as an AVEC Sub-challenge in 2018~\cite{Ringeval18-A2W}. Whereas the AVEC 2018 CES focused on detecting \emph{arousal}, \emph{valence}, and \emph{liking} from Hungarian speakers using only German speakers for training and development of the models~\cite{Ringeval18-A2W}, in this year's AVEC CES the test cohort is Chinese speakers with speakers from the two cultures mentioned earlier being available for training, development, and additional testing. 

A common belief in facial expression recognition is that emotional expressions have a large degree of universality across cultures~\cite{Corneanu16-SOR, Ekamn71-UAC}. This statement was on the whole supported by both baseline results and works submitted to the AVEC 2018 CES, with either vision-only or multimodal systems achieving higher cross-culture accuracies than speech-only approaches~\cite{Ringeval17-SFA, Huang18-MCE, Wataraka18-SBC, Zhao18-MMD}. These results were insightful, as previously, there were only a few works in the affective computing literature which supported this claim~\cite{cordaro2018universals, Dmello15-ARA}.

Interestingly, approaches in the AVEC 2018 CES did not employ approaches such as transfer learning~\cite{Zhang17-EDE, Zhang17-CCA} or domain adaptation techniques~\cite{Kaya18-EAE, Sagha16-CLS} typically seen in cross-cultural testing. In~\cite{Wataraka18-SBC}, the authors proposed a model based  on emotional salient detection to identify emotion markers invariant to socio-cultural context. The other two entrants employed data driven approaches based on long short-term memory recurrent neural networks (LSTM-RNN)~\cite{Huang18-MCE, Zhao18-MMD}. Matching with similar results in the literature~\cite{Feraru15-CAE, Scherer01-EIF}, all entrants in the AVEC 2018 CES observed a drop in system performance when testing on the Hungarian data~\cite{Huang18-MCE, Wataraka18-SBC, Zhao18-MMD}.
\section{Challenge corpora}
\label{Challenge_corpora}
The AVEC 2019 Challenge relies on three corpora: (i) the USoM corpus~\cite{rathner18som} for the SoMS, (ii) the Extended-DAIC corpus~\cite{gratchetal14} for the DDS, and (iii) the SEWA dataset~\cite{kossaifi19-sda} for the CES. We provide below a short overview of each dataset and refer the reader to the original work for a more complete description.

\subsection{Ulm State-of-Mind Corpus}

The Ulm state of mind database was collected to assess the association between personal story telling and current SOM, operationalised by affective state according to Russel's theory~\cite{schuller2018interspeech,rathner18som,russell2003core}. 
Parts of this dataset have been released for the Interspeech 2018 Computational Paralinguistics (ComParE) challenge \cite{schuller2018interspeech}. 

Participants of the USoM corpus were instructed to first tell two negative personal narratives $NN_{1,2}$ and subsequently two positive personal narratives $PN_{1,2}$, each for five minutes in front of a camera. They were also asked to rate their current affect ($CA$) on a 10-point likert scale for the dimensions \textit{arousal} and \textit{valence} before and after telling each narrative, resulting in the following protocol: $(t_0)$, $CA_0, NN_1$, $(t_1)$, $CA_1, NN_2$, $(t_2)$, $CA_2, PN_1$, $(t_3)$, $CA_3, PN_2$, and $(t_4)$, $CA_4$. 
For the purpose of the Challenge, the USoM dataset was partitioned into training, development, and test sets while preserving the overall speaker diversity -- in terms of age, gender distribution, and core affect evaluations -- within the partitions. Table~\ref{tab:usom} shows the number of subjects and duration for each partition. 

As the interest of the SoMS is on the change in mood, rather than just its static observation, the initial self-reports made before the storytelling are included in the data package given to participants for all partitions, including the test set. Exploiting such contextual information in an automatic system predicting the level of mood is a realistic scenario in the real-world, because a therapist would always ask a person's baseline emotion at the start of a session. It is thus essential to provide machine learning algorithms with the same prior information as a therapist would have.


\subsection{Distress Analysis Interview Corpus}

The Extended Distress Analysis Interview Corpus (E-DAIC)~\cite{DeVaultEtAl2014_SVH} is an extended version of WOZ-DAIC \cite{gratchetal14} that contains semi-clinical interviews designed to support the diagnosis of psychological distress conditions such as anxiety, depression, and post-traumatic stress disorder. These interviews were collected as part of a large effort to create a computer agent that interviews people and identifies verbal and nonverbal indicators of mental illnesses~\cite{gratchetal14}. 

Data collected include audio and video recordings, automatically transcribed text using Google Cloud's speech recognition service, and extensive questionnaire responses. The interviews are conducted by an animated virtual interviewer called Ellie. In the WoZ interviews, the virtual agent is controlled by a human interviewer (wizard) in another room, whereas in the AI interviews, the agent acts in a fully autonomous way using different automated perception and behaviour generation modules. 

For the purpose of the Challenge, the E-DAIC dataset was partitioned into training, development, and test sets while preserving the overall speaker diversity -- in terms of age, gender distribution, and the eight-item Patient Health Questionnaire (PHQ-8) scores -- within the partitions. Whereas the training and development sets include a mix of WoZ and AI scenarios, the test set is solely constituted from the data collected by the autonomous AI. Details regarding the speaker distribution over the partitions are given in Table~\ref{tab:e-daic}.

\begin{table}[t]
  \caption{Number of subjects and duration of the storytellings contained in the USoM database~\cite{rathner18som}.}
  \label{tab:usom}
  \begin{tabular}{lcr}
    \toprule
    \textbf{Partition} & \textbf{\# Subjects} & \textbf{Duration [h:min:s]} \\
    \midrule
    Training    &  45 &   13:49:38 \\
    Development &  33 &   10:46:57 \\
    Test        &  33 &   9:46:14 \\
    \midrule
    \textbf{All} & \textbf{111} & \textbf{34:22:49} \\
    \bottomrule
\end{tabular}
\end{table}

\begin{table}[t]
  \caption{Number of subjects and duration of the interviews included in the Extended-DAIC database~\cite{gratchetal14}.}
  \label{tab:e-daic}
  \begin{tabular}{lcr}
    \toprule
    \textbf{Partition} & \textbf{\# Subjects} & \textbf{Duration [h:min:s]} \\
    \midrule
    Training    &  163 &   43:30:20 \\
    Development &  56 &   14:47:31 \\
    Test        &  56 &   14:52:42 \\
    \midrule
    \textbf{All} & \textbf{275} & \textbf{73:10:33} \\
    \bottomrule
\end{tabular}
\end{table}

\begin{table}[t]
  \caption{Number of subjects and duration of the video chats contained in the SEWA database~\cite{kossaifi19-sda}.}
  \label{tab:sewa}
  \begin{tabular}{llcr}
    \toprule
    \textbf{Culture} & \textbf{Partition} & \textbf{\# Subjects} & \textbf{Duration [h:min:s]} \\
    \midrule
    German    & Training    &  34 &   1:33:12 \\
    German    & Devel. &  14 &   0:37:46 \\
    German    & Test        &  16 &   0:46:38 \\
    Hungarian & Training    &  34 &   1:08:24 \\
    Hungarian & Devel. &  14 &   0:28:42 \\
    Hungarian & Test        &  18 &   0:36:06 \\
    Chinese   & Test        &  70 &   3:17:52 \\
    \midrule
    \textbf{All} &  & \textbf{200} & \textbf{8:28:40} \\
    \bottomrule
\end{tabular}
\end{table}

\subsection{Cross-cultural Emotion Database (SEWA)}

The SEWA database consists of audiovisual recordings of spontaneous behaviour of participants captured using an \emph{in-the-wild} recording paradigm~\cite{kossaifi19-sda}. Pairs of friends or relatives from German, Hungarian, and Chinese cultures were recorded through a dedicated video chat platform which utilised participants' own -- standard -- web-cameras and microphones. After watching a set of commercials, pairs of participants were given the task of discussing the last advert watched (a video clip advertising a water tap) for up to three minutes. The aim of this discussion was to elicit further reactions and opinions about the advert and the product advertised. 

The video chats of the three cultures have been annotated w.\,r.\,t.\ the emotional dimensions \emph{arousal} and \emph{valence}, and a third dimension describing \emph{liking} (or sentiment), independently by several native speakers; German and Chinese: six annotators, Hungarian: five annotators. The annotation contours (traces) are combined into a single gold-standard using the same \emph{evaluator weighted estimator (EWE)}-based approach that was used in the last two editions of AVEC~\cite{ringeval17a2r, Ringeval18-A2W}. Table~\ref{tab:sewa} shows the number of subjects and the duration of the recordings for each partition.

\section{Baseline features}
\label{Baseline_features}
Emotion recognition from audiovisual signals usually relies on feature sets whose extraction is based on expertise gained over several decades of research in the domains of speech processing, e.\,g., Mel Frequency Cepstral Coefficients (MFCCs), and vision computing, e.\,g., Facial Action Units (FAUs). However, recent advances in the field of representation learning, whose objective is to learn representations of data that are best suited for the recognition task~\cite{Bengio13-MMT}, have shown that efficient representations of audiovisual signals can be learnt in the context of emotion~\cite{Trigeorgis16-AFE,Schmitt16-ATB,Amiriparian17-SSC}. 

Audiovisual representations can be learnt from expert-driven information extracted from the raw signals~\cite{Schmitt16-ATB}, or directly from the raw signals~\cite{Trigeorgis16-AFE}. They can also be generated using adversarial networks~\cite{Deng17-SDO}, or using convolutional neural networks trained on out-of-domain data and for a different task, \eg audio representations extracted by a model trained for object classification in images~\cite{Amiriparian17-SSC}.

\subsection{Expert-knowledge}
The traditional approach in affect sensing consists in summarising low-level descriptors (LLDs) of audiovisual signals over time with a set of statistical measures computed over a fixed-duration sliding analysis window. Those descriptors usually include spectral, cepstral, prosodic, and voice quality information for the audio channel, and appearance, geometric, and FAUs information for the video channel. 

As audio features, we compute the extended Geneva Minimalistic Acoustic Parameter Set (\textsc{eGeMAPS})~\cite{Eyben16-TGM}, which contains 88 measures covering the aforementioned acoustic dimensions, and used here as baseline. In addition, MFCCs 1-13, including their 1$^{\mathrm{st}}$- and 2$^{\mathrm{nd}}$-order derivatives (deltas and double-deltas) are computed as a set of acoustic LLDs, using the \textsc{openSMILE\footnote{\url{http://audeering.com/technology/opensmile/}}}~\cite{Eyben13-RDI} toolkit. As visual features, we extract the intensities of 17 FAUs for each video frame, along with a confidence measure, using the toolkit \textsc{openFace}\footnote{\url{https://github.com/TadasBaltrusaitis/OpenFace/}}~\cite{baltrusaitis2018openface}. Descriptors of pose and gaze are additionally extracted. 

Audiovisual LLDs are summarised over time by computing their mean and standard-deviation using a sliding window of 4\,s length, and a hop size of 1\,s for the USoM and E-DAIC datasets, and 100\,ms for the SEWA dataset, excepted for the \textsc{eGeMAPS} set, which is computed on each window.

\subsection{Bags-of-Words}

The technique of bags-of-words (BoW), which originates from text processing, represents the distribution of LLDs according to a dictionary learnt from them. As a front-end of the BoW, we use the MFCCs and the \textsc{eGeMAPS} set for the acoustic data, and the intensities of the FAUs for the video data; MFCCs and \textsc{eGeMAPS} LLDs are standardised (zero mean, unit variance) in an on-line approach prior to vector quantisation, while this step is not required for the FAU intensities. 

To generate the BoW representations, both the acoustic and the visual features are processed and summarised over a block of a 4\,s length duration, for each step of 100\,ms for the SEWA dataset, and 1\,s for the USoM and E-DAIC datasets. 
The codebook size is $100$. 
Instances are sampled at random to build the dictionary, and the logarithm is taken from resulting term frequencies in order to compress their range. The whole cross-modal BoW (XBoW) processing chain is executed using the open-source toolkit \textsc{openXBOW}\footnote{\url{https://github.com/openXBOW/openXBOW}}~\cite{Schmitt17-OIT}.

\subsection{Deep Representations}

As in last year's challenge~\cite{Ringeval18-A2W}, we have included \textsc{Deep Spectrum}\footnote{\url{https://github.com/DeepSpectrum/DeepSpectrum}} features as a deep learning based audio baseline feature representation~\cite{Amiriparian17-SSC}. \textsc{Deep Spectrum} features are inspired by deep representation learning paradigms common in image processing: spectral images of speech instances are fed into pre-trained image recognition CNNs and a set of the resulting activations are extracted as feature vectors. 

For this year's challenge, we extracted \textsc{Deep Spectrum} features from four robust pre-trained CNNs  using \textsc{VGG-16}~\cite{simonyan2014very}, \textsc{AlexNet}~\cite{Krizhevsky2012}, \textsc{DenseNet-121}, and \textsc{DenseNet-201}~\cite{Huang_2017_CVPR}; \textsc{AlexNet} was used in the AVEC 2019 CES purely for consistency with the previous AVEC 2018 CES. The speech files are first transformed into mel-spectrogram images with 128 mel-frequency bands, a window width of 4\,s for all challenge corpora and a hop size of 1\,s for the USoM and E-DAIC datasets, and 100\,ms for the SEWA dataset. Following that, the spectral-based images are forwarded through the pre-trained networks. A  4\,096-dimensional feature vector is then formed from the activations of the second fully connected layer in \textsc{VGG-16} and \textsc{AlexNet}, and a 1\,024 and a 1\,920-dimensional feature vector is obtained from the activations of the last average pooling layer of the \textsc{DenseNet-121} and \textsc{DenseNet-201} networks, respectively.


We also provide two baseline deep visual representations. For these, we employed a VGG-16 \cite{simonyan2014very} network and a ResNet-50 network \cite{he2016deep} that are pre-trained with the Affwild dataset \cite{kollias2019deep}. The pipeline starts with applying the \textsc{openFace} toolkit~\cite{baltrusaitis2018openface} to detect the face region and subsequently perform face alignment. Then, we froze the weights of two pre-trained models and fed the aligned face images to both CNNs individually. To obtain the deep representations for each frame, we extract the output of the first fully-connected layer from the pre-trained VGG-16 network, and the output of the global average pooling layer from the pre-trained ResNet-50 network, respectively. As a result, a 4\,096-dimensional deep feature vector from VGG and a 2\,048-dimensional deep feature vector from ResNet are provided for each frame.   

\begin{table*}
  \caption{Baseline results evaluated with $CCC$ for the AVEC 2019 SoMS; USoM data set~\cite{rathner18som}; BoAW-M/e: bags-of-audio-words with MFCCs/eGeMAPS; DS-DNet: Deep Spectrum using DenseNet-121; DS-VGG: Deep Spectrum using VGG-16; best result on the test partition is highlighted in bold.}
  \label{tab:baseline_soms}
  \begin{tabular}{lcccccccccccc}
    \toprule
    &\multicolumn{6}{c}{\textbf{\rule[2 pt]{105 pt}{0.5 pt} Audio \rule[2 pt]{105 pt}{0.5 pt}}} & \multicolumn{4}{c}{\textbf{\rule[2 pt]{45 pt}{0.5 pt} Video \rule[2 pt]{45 pt}{0.5 pt}}} &\textbf{Fusion}\\
    \midrule
    \textbf{Partition}&MFCCs	&eGeMAPS &BoAW-M &BoAW-e &DS-DNet &DS-VGG &FAUs 	&BoVW &ResNet &VGG &\textit{All}\\
    \midrule
    &\multicolumn{11}{c}{\textit{Random sampling of training instances}}\\
    \midrule
    Development &.282  &.412  &.336  &.295 &.280 &.384 &.372 &.317 &.261 &.318 &.417\\
    Test &--  &.276  &--  &-- &-- &\textbf{.289} &.119 &-- &-- &.191 &.278\\
    \midrule
    &\multicolumn{11}{c}{\textit{Curriculum sampling of training instances}}\\
    \midrule
    Development &.299  &.378  &.334  &.288 &.326 &.437 &.419 &.313 &.300 &.318 &.464\\
    Test &--  &\textbf{.294}  &--  &-- &-- &.208 &.151 &-- &- &.160 &.236\\
    \bottomrule
\end{tabular}
\end{table*}

\section{Baseline systems}
\label{Baseline_systems}

All baseline systems rely exclusively on existing open-source machine learning toolkits to ensure the reproducibility of the results. In this section, we describe the systems developed for each Sub-challenge, and present the obtained results. For evaluation on the test set, we retained the two audio representations with the best performance, and the two video representations with the best performance, in addition to the fusion of all audiovisual representations.

\subsection{State-of-Mind Sub-challenge}

We use a gated recurrent unit (GRU) network with two layers, each having 64 nodes for their hidden layers, for each audiovisual representation. As a pre-processing step, all input features are normalised to have zero mean and unit variance. Dropout, at a rate of 10\,\%, is employed during training. The GRU is then followed by a fully connected neural network that has one hidden layer with 32 nodes, followed by a single linear layer to map to the desired output size of one. Note that a middle-fusion of the audiovisual representations is performed by concatenating their respective GRU outputs.

The model is implemented using a \textsc{Pytorch} framework and is trained with an \textsc{Adam} optimiser. As previous studies have shown the benefits of training a network following a curriculum~\cite{Bengio09-CL,Lotfian19-CLF}, 
where instances are gradually presented in increasing level of difficulty, we implemented this approach using the following strategy. 
First, a uniform distribution of valence labels is obtained by duplicating training instances, then, a sub-set of the training set with only the data instances with $CA \in [2 - 3] \cup [9 - 10]$, \ie the most positive and negative storytellings, is firstly used for training, followed by a larger sub-set with data instances with $CA \in[2 - 4] \cup [8 - 10]$, each for 32 epochs. We then exploited the whole training set until \textit{early stopping} occurs; once 60 epochs have passed, training is stopped if there is no improvement within the last 25 epochs. 

Because the interest of the SoMS is in the analysis of a change in human SOM, the network is trained to model the difference between the self-reported core affect after each story and before the first story: $CA_i - CA_0, i=1,2,3,4$. Results are reported for each audiovisual representation, and for the two training approaches, \ie with or without curriculum, in Table~\ref{tab:baseline_soms}. Whereas the mid-fusion of all audiovisual representations provides the best result on the development set for the two learning approaches, audio descriptors achieve higher performance on the test set, with the expert-based eGeMAPS set performing best with curriculum learning. 

A summary of the results obtained with either a static ($CA_i$) or a dynamic ($CA_i - CA_0$) view of the self-reported mood used for training or testing the system is also provided in Table~\ref{tab:baseline_soms_summary}. 
Interestingly, results show that the automatic inference of the self-reported mood performs much better in a `mixed' scenario, \ie training on the static view ($CA_i$) and evaluating on the change ($CA_i - CA_0$) or \textit{vice-versa} training on the change and testing on the static label, compared to a `consistent' approach with both training and testing performed on the same view, \ie either static or dynamic.

This result might stem from emotion data being hierarchically organised. As such, each self-reported emotion is nested within a person over a period of time~\cite{koval2012getting}. Because of human's inability to assess their own emotions as an absolute value, self-reported emotion can only be interpreted as a current assessment of emotional differences in relation to the nearest past. Furthermore, there is  also variance in emotion dynamics between people and not only within a person~\cite{koval2013affect}. The inter-individual and intra-individual variance in emotion dynamics are strongly related to one another, but add both new information to predictions. While the variance between persons might be best captured in a scenario where machine learning is applied to raw values, the intra-individual auto-correlation of emotion, the so-called emotional inertia, is portrayed in the dynamic evaluation~\cite{kuppens2010emotional}. Therefore, training on static data and evaluating on dynamic data, such as emotional inertia, might be the \textit{state-of-the-art} approach to characterise human SOM. 

\begin{table}
  \caption{Comparison of the approaches -- training or testing on a static or dynamic measure of mood -- used for the AVEC 2019 SoMS; averaged $CCC$ results are reported; $[\mu (\sigma)]$.}
  \label{tab:baseline_soms_summary}
  \begin{tabular}{lcc}
    \toprule
    \textbf{Partition} &Static training	&Dynamic training \\
    \midrule
    &\multicolumn{2}{c}{\textit{Static evaluation}}\\
    \midrule
    Development &.149 (.108) &.335 (.050)\\
    Test &.037 (.063) &.219 (.068)\\
    \midrule
    &\multicolumn{2}{c}{\textit{Dynamic evaluation}}\\
    \midrule
    Development &.368 (.150) &.102 (.066)\\
    Test &\textbf{.325 (.052)} &.040 (.094)\\
    \bottomrule
\end{tabular}
\end{table}

\begin{table*}
  \caption{Baseline results evaluated with $CCC$ for the AVEC 2019 DDS; $RMSE$ is additionally reported; BoAW-M/e: bags-of-audio-words with MFCCs/eGeMAPS; DS-DNet: Deep Spectrum using DenseNet-121; DS-VGG: Deep Spectrum using VGG-16; best result on the test partition is highlighted in bold.}
  \label{tab:baseline_dds}
  \begin{tabular}{lccccccccccc}
    \toprule
    &\multicolumn{6}{c}{\textbf{\rule[2 pt]{105 pt}{0.5 pt} Audio \rule[2 pt]{105 pt}{0.5 pt}}} & \multicolumn{4}{c}{\textbf{\rule[2 pt]{45 pt}{0.5 pt} Video \rule[2 pt]{45 pt}{0.5 pt}}} &\textbf{Fusion}\\
    \midrule
    \textbf{Partition}&MFCCs	&eGeMAPS &BoAW-M &BoAW-e &DS-DNet &DS-VGG &FAUs 	&BoVW &ResNet &VGG &\textit{All}\\
    \midrule
    &\multicolumn{11}{c}{\textit{Regression of PHQ-8 score ($CCC$)}}\\
    \midrule
    Development & .198 & .076 & .102 & .272 &.165 &.305 &.115 & .107 & .269 & .108 &.336\\
    Test &-- &-- &-- &.045 &-- &.108 & .019 &-- &\textbf{.120} &-- &.111\\
    \midrule
    &\multicolumn{11}{c}{\textit{Regression of PHQ-8 score ($RMSE$)}}\\
    \midrule
    Development & 7.28  & 7.78 & 6.32 &6.43 &8.09 &8.00 &7.02 &5.99 &7.72 &7.69 &5.03\\
    Test &-- &-- &-- &8.19 &-- &9.33 &10.0 &-- &8.01 &-- &\textbf{6.37}\\
    \bottomrule
\end{tabular}
\end{table*}

\subsection{Detecting Depression Sub-challenge}
For the depression detection baseline, we employ a single-layer 64-d GRU as our recurrent network with a dropout regularisation of rate 20\,\%, followed by a 64-d fully-connected layer to obtain a single-value regression score. To handle bias, we convert the PHQ-8 score labels to floating point numbers by downscaling with a factor of 25 prior to training. The network is trained and evaluated using a $CCC$ loss function and evaluation score, and the $RMSE$ results are reported using the original PHQ scale. A batch size of 15 is used consistently, and the learning rate is optimised across different feature sets. In order for the data to fit on GPU memory, a maximum sequence length has been assigned for the sessions. For the MFCCs and eGeMAPS LLDs, and the high dimensional deep representations like DeepSpectrum, ResNet, and VGG, a maximum sequence length of 20\,minutes is used. Additionally, for ResNet, VGG, and \textsc{Deep Spectrum} representations frames are dropped keeping one out of two, or one out of four frames depending on the dimensionality so that the data can be loaded onto memory. Fusion of the different audiovisual representations is achieved by 
averaging their scores.


Baseline results of the DDS are given in Table~\ref{tab:baseline_dds}. They show that, on the development set, the best $CCC$ score from audio features was achieved with Deep spectrum (DS-VGG) features, and the model with ResNet features achieved the best result for visual features. 
These results indicate the power of representations learnt by deep neural networks with a large amount of data when being used in a different context to which they were initially designed, which is confirmed on the test set with the ResNet visual model achieving the best result, despite a relatively low $CCC$. 

Fusion of the different representations achieves the best result on the development set, and the $RMSE$ returned on the test set is slightly better than the one obtained on the DAIC-WoZ dataset with the AVEC 2017 baseline system~\cite{ringeval17a2r}; $RMSE=6.37$ for AVEC 2019 compared to $RMSE=6.97$ for AVEC 2017. However, the baseline system developed for this year's Challenge is more complex -- a simple linear regression model \textit{vs} GRU-RNNs for this year --, and the corresponding scores should be therefore best regarded in the light of the best results of the AVEC 2017 Depression Sub-challenge~\cite{Gong17-TMB}, which was $RMSE=4.99$. 

On the basis of the results obtained in the automatic sensing of the level of depression from interactions with the virtual agent, recognition seems more challenging when the agent is solely AI driven, than when a human is driving the agent as a WoZ.
This observation opens interesting research questions for designing the agent in a way that the observation of depression cues can be maximised, \eg by reinforcement learning, according to the interaction style of the agent.

\begin{table*}
  \caption{Baseline results evaluated with $CCC$ for the AVEC 2019 CES; SEWA dataset~\cite{kossaifi19-sda}; DeepSpec: Deep Spectrum; best result on the test partition is highlighted in bold.}
  \label{tab:baseline_ces}
  \begin{tabular}{lcccccccccccc}
    \toprule
    & &\multicolumn{5}{c}{\textbf{\rule[2 pt]{75 pt}{0.5 pt} Audio \rule[2 pt]{75 pt}{0.5 pt}}} & \multicolumn{4}{c}{\textbf{\rule[2 pt]{45 pt}{0.5 pt} Video \rule[2 pt]{45 pt}{0.5 pt}}} &\textbf{Fusion}\\
    \midrule
    \textbf{Culture} &\textbf{Partition} &MFCCs	&eGeMAPS &BoAW-M &BoAW-e &DS &FAUs 	&BoVW &ResNet &VGG &\textit{All}\\
    \midrule
    & &\multicolumn{10}{c}{\textit{Arousal}}\\
    \midrule
    German &Dev.    &.389  &.396  &.323  &.434 &.380 &.606 &.556 &.475 &.561 &.629\\
    German &Test    &--   &.293  &-- 	&.276  &-- &.562 &-- &-- &.505 &.517\\
    Hungarian &Dev.    &.236 &.305  &.237 	&.291  &.156 &.425 &.321 &.460 &.367 &.583\\
    Hungarian &Test    &--   &.272  &-- 	&.250  &-- &.527 &-- &-- &.396 &.525\\
    Ger. + Hun. &Dev.    &.326   &.371  &.298 	&.398  &.312 &.531 &.467 &.473 &.493 &.614\\
    Chinese &Test    &--   &.100  &-- 	&.107  &-- &\textbf{.355} &-- &-- &.297 &.238\\
    \midrule
    & &\multicolumn{10}{c}{\textit{Valence}}\\
    \midrule
    German &Dev.   &.344  &.405  &.190  &.455 &.317 &.639 &.594 &.552 &.595 &.684\\
    German &Test   &--  &.309  &--  &.325 &-- &.627 &-- &-- &.646 &.622\\
    Hungarian &Dev.   &.017   &.073  &.042 	&.135  &.084 &.463 &.421 &.373 &.363 &.508\\
    Hungarian &Test   &--   &.166  &-- 	&.151  &.173 &.459 &-- &-- &.548 &.397\\
    Ger. + Hun. &Dev.    &.187   &.286  &.134 	&.352  &.233 &.565 &.523 &.487 &.505 &.615\\
    Chinese &Test    &.-- &.267 &-- &.281 &-- &\textbf{.468} &-- &-- &.398 &.423\\
    \midrule
    & &\multicolumn{10}{c}{\textit{Liking}}\\
    \midrule
    German &Dev.   &.159  &.136  &.140  &.003 &.164 &.056 &.073 &.057 &.244 &.048\\
    German &Test   &--  &.012  &--  &.074 &-- &-.042 &-- &-- &-.052 &-.019\\
    Hungarian &Dev.   &.115   &.192  &-.027 	&.253  &.121 &.104 &.041 &.028 &.028 &.260\\
    Hungarian &Test   &--   &.051  &-- 	&.089  &-- &-.062 &-- &-- &-.069 &-.22\\
    Ger. + Hun. &Dev.    &.144   &.159  &.074 	&.138  &.142 &.083 &.057 &.040 &.037 &.222\\
    Chinese &Test    &--   &.007  &-- &\textbf{.041} &-- &.006 &-- &-- &-.006 &-.012\\
    \bottomrule
\end{tabular}
\end{table*}
\subsection{Cross-cultural Emotion Sub-challenge}

For the baseline system of the CES, we employ a 2-layer LSTM-RNN (64 / 32 units) as a time-dependent regressor of the three targets (learnt together) for each representation of the audiovisual signals, and SVMs -- \textsc{liblinear} with L2-L2 dual form of the objective function -- for the late fusion of the predictions. The model is implemented using the \textsc{Keras} framework. The network is trained for 50 epochs with the \textsc{RMSprop} optimiser using a dropout rate of 10\,\%, and the model providing the highest CCC on the development set of the German and Hungarian culture is used to generate the predictions for the test sets (German, Hungarian, and all clips of the Chinese culture). Even though the model has three outputs modelling each dimension, the optimum model for each dimension is selected separately. The predictions of all test sequences from each culture are concatenated prior to computing the $CCC$, whose opposite is used as loss function for training the networks~\cite{Trigeorgis16-AFE,Weninger16-DTR}. 

In order to perform time-continuous prediction of the emotional dimensions, audiovisual signals were processed with a sliding window of 4\,s length, which is a compromise to capture enough information to be used with both static regressors, such as SVMs, and context-aware regressors, such as RNNs. We utilised frame-stacking for the SVM-based late fusion of the audiovisual representations with either past, or future context. 





Baseline results of the CES are given in Table~\ref{tab:baseline_ces}. They show improvements over the performance reported in the previous edition of the AVEC CES; relative improvement for German is 7.25\% and 8.25\% for arousal and valence, respectively, and for Hungarian, 17.3\% and 13.3\%, respectively. The inclusion of instances of the Hungarian culture as training and development material, in addition to those of the German culture, might explain the large increase in performance for both cultures, as only instances of the German culture were available for training and development in the AVEC 2018 CES. In addition, a more recent version of the \textsc{openFace} toolkit~\cite{baltrusaitis2018openface} was exploited, which provided the best results on the test set for both arousal and valence with FAUs based features. Those results confirm the common view that facial expressions of emotion have a large degree of universality across cultures, compared to the vocal expressions, where the acoustic and prosodic dimensions already play a key role in the oral communication by serving many grammatical and pragmatic functionalities, \eg in tonal languages like Mandarin, the meaning associated with a syllable depends on its pitch contour. Such language dependent peculiarities make cross-cultural settings highly challenging, especially when noise comes into play because of the ecological conditions of study.


\section{Conclusions}
\label{Conclusion}

In this paper, we introduced AVEC 2019 -- the sixth combined open Audio/Visual Emotion and Health assessment challenge. It comprises three Sub-challenges: i) SoMS, where the level of mood has to be predicted from positive and negative personal stories; ii) DDS, where the level of depression (PHQ-8 score) has to be predicted from structured interviews conducted by a virtual agent wholly driven by AI; and, iii) CES, where the level of affective dimensions of \emph{arousal}, \emph{valence}, and \emph{liking} has to be inferred in a cross-cultural \emph{in-the-wild} paradigm with German and Hungarian cultures as training and testing material, and Chinese culture as solely testing material. 

By intention, we opted to use exclusively open-source software and the highest possible transparency and realism for the baselines, by using the same number of trials as given to participants for reporting results on the test partition, and sharing all the developed scripts for both features extraction and machine learning on a public platform. Results indicate that: i) in the SoMS, the level of mood was best predicted when the system was trained on the static scores and evaluated on their dynamic view, \ie between the label provided after the storytellings, and before the first story, which can be explained by inertial emotion theories; ii) in the DDS, prediction of the level of depression (PHQ-8) is reported to be more challenging when the virtual agent conducting the interview is wholly driven by AI, compared to a WoZ setup; and iii), in the CES, dimensional emotions are more challenging to sense in a cross-cultural setting for audio descriptors compared to video descriptors, which confirm on one hand the universality of facial expressions for Asian (Chinese) and Western European cultures (German and Hungarian), and show on the other the challenge of using audio descriptors for paralinguistics analysis in languages presenting dissimilarities in their acoustic, in particular when data are collected in an ecological (noisy) environment. 



\begin{acks}
The research leading to these results has received funding from the Horizon 2020 Programme through 
the Research Innovation Action No. 826506 (sustAGE), and No.\ 688835 (DE-ENIGMA). Further funding has also been received from the Innovative Medicines Initiative 2 Joint Undertaking under grant agreement No.\,115902, which receives support from the European Union's Horizon 2020 research and innovation program and EFPIA. The work on the DDS was supported in part by the U.S. Army. Any opinion, content or information presented does not necessarily reflect the position or the policy of the United States Government, and no official endorsement should be inferred. The authors further thank the sponsor of the challenge -- audEERING GmbH. 

\end{acks}
 
\bibliographystyle{ACM-Reference-Format}
\balance
\bibliography{biblio}


\begin{thebibliography}{82}


\ifx \showCODEN    \undefined \def \showCODEN     #1{\unskip}     \fi
\ifx \showDOI      \undefined \def \showDOI       #1{#1}\fi
\ifx \showISBNx    \undefined \def \showISBNx     #1{\unskip}     \fi
\ifx \showISBNxiii \undefined \def \showISBNxiii  #1{\unskip}     \fi
\ifx \showISSN     \undefined \def \showISSN      #1{\unskip}     \fi
\ifx \showLCCN     \undefined \def \showLCCN      #1{\unskip}     \fi
\ifx \shownote     \undefined \def \shownote      #1{#1}          \fi
\ifx \showarticletitle \undefined \def \showarticletitle #1{#1}   \fi
\ifx \showURL      \undefined \def \showURL       {\relax}        \fi
\providecommand\bibfield[2]{#2}
\providecommand\bibinfo[2]{#2}
\providecommand\natexlab[1]{#1}
\providecommand\showeprint[2][]{arXiv:#2}

\bibitem[\protect\citeauthoryear{Althoff, Clark, and Leskovec}{Althoff
  et~al\mbox{.}}{2016}]%
        {althoff2016large}
\bibfield{author}{\bibinfo{person}{Tim Althoff}, \bibinfo{person}{Kevin Clark},
  {and} \bibinfo{person}{Jure Leskovec}.} \bibinfo{year}{2016}\natexlab{}.
\newblock \showarticletitle{{Large-scale Analysis of Counseling Conversations:
  An Application of Natural Language Processing to Mental Health}}.
\newblock \bibinfo{journal}{\emph{Transactions of the Association for
  Computational Linguistics}}  \bibinfo{volume}{4} (\bibinfo{year}{2016}),
  \bibinfo{pages}{463--476}.
\newblock


\bibitem[\protect\citeauthoryear{Amiriparian, Gerczuk, Ottl, Cummins, Freitag,
  Pugachevskiy, Baird, and Schuller}{Amiriparian et~al\mbox{.}}{2017}]%
        {Amiriparian17-SSC}
\bibfield{author}{\bibinfo{person}{Shahin Amiriparian},
  \bibinfo{person}{Maurice Gerczuk}, \bibinfo{person}{Sandra Ottl},
  \bibinfo{person}{Nicholas Cummins}, \bibinfo{person}{Michael Freitag},
  \bibinfo{person}{Sergey Pugachevskiy}, \bibinfo{person}{Alice Baird}, {and}
  \bibinfo{person}{Bj\"orn Schuller}.} \bibinfo{year}{2017}\natexlab{}.
\newblock \showarticletitle{{Snore sound classification using image-based deep
  spectrum features}}. In \bibinfo{booktitle}{\emph{{Proc.\ of INTERSPEECH
  2017, 18th Annual Conference of the International Speech Communication
  Association}}}. \bibinfo{publisher}{ISCA}, \bibinfo{address}{Stockholm,
  Sweden}, \bibinfo{pages}{3512--3516}.
\newblock


\bibitem[\protect\citeauthoryear{Association}{Association}{2013}]%
        {APA13-DSM}
\bibfield{author}{\bibinfo{person}{American~Psychiatric Association}.}
  \bibinfo{year}{2013}\natexlab{}.
\newblock \bibinfo{booktitle}{\emph{{Diagnostic and Statistical Manual of
  Mental Disorders (DSM-5)}}}.
\newblock \bibinfo{publisher}{American Psychiatric Publishing},
  \bibinfo{address}{Arlington, VA}.
\newblock


\bibitem[\protect\citeauthoryear{Baltru\v{s}aitis, Zadeh, Lim, and
  Morency}{Baltru\v{s}aitis et~al\mbox{.}}{2018}]%
        {baltrusaitis2018openface}
\bibfield{author}{\bibinfo{person}{Tadas Baltru\v{s}aitis},
  \bibinfo{person}{Amir Zadeh}, \bibinfo{person}{Yao~Chong Lim}, {and}
  \bibinfo{person}{Louis-Philippe Morency}.} \bibinfo{year}{2018}\natexlab{}.
\newblock \showarticletitle{{OpenFace 2.0: Facial Behavior Analysis Toolkit}}.
  In \bibinfo{booktitle}{\emph{Proc.\ 13th IEEE International Conference on
  Automatic Face \& Gesture Recognition (FG 2018)}}. \bibinfo{publisher}{IEEE},
  \bibinfo{address}{Xi'an, P.\,R.\ China}, \bibinfo{pages}{59--66}.
\newblock


\bibitem[\protect\citeauthoryear{Baumel and Yom-Tov}{Baumel and
  Yom-Tov}{2018}]%
        {baumel2018predicting}
\bibfield{author}{\bibinfo{person}{Amit Baumel} {and} \bibinfo{person}{Elad
  Yom-Tov}.} \bibinfo{year}{2018}\natexlab{}.
\newblock \showarticletitle{Predicting user adherence to behavioral eHealth
  interventions in the real world: examining which aspects of intervention
  design matter most}.
\newblock \bibinfo{journal}{\emph{Translational Behavioral Medicine}}
  \bibinfo{volume}{8}, \bibinfo{number}{5} (\bibinfo{year}{2018}),
  \bibinfo{pages}{793--798}.
\newblock


\bibitem[\protect\citeauthoryear{Bengio, Courville, and Vincent}{Bengio
  et~al\mbox{.}}{2013}]%
        {Bengio13-MMT}
\bibfield{author}{\bibinfo{person}{Yoshua Bengio}, \bibinfo{person}{Aaron
  Courville}, {and} \bibinfo{person}{Pascal Vincent}.}
  \bibinfo{year}{2013}\natexlab{}.
\newblock \showarticletitle{{Representation Learning: A Review and New
  Perspectives}}.
\newblock \bibinfo{journal}{\emph{{IEEE Transactions on Pattern Analysis and
  Machine Intelligence}}} \bibinfo{volume}{35}, \bibinfo{number}{4}
  (\bibinfo{date}{August} \bibinfo{year}{2013}), \bibinfo{pages}{1798--1828}.
\newblock


\bibitem[\protect\citeauthoryear{Bengio, Louradour, Collobert, and
  Weston}{Bengio et~al\mbox{.}}{2009}]%
        {Bengio09-CL}
\bibfield{author}{\bibinfo{person}{Yoshua Bengio}, \bibinfo{person}{J\'er\^ome
  Louradour}, \bibinfo{person}{Ronan Collobert}, {and} \bibinfo{person}{Jason
  Weston}.} \bibinfo{year}{2009}\natexlab{}.
\newblock \showarticletitle{{Curriculum Learning}}. In
  \bibinfo{booktitle}{\emph{Proc.\ International Conference on Machine Learning
  (ICML)}}. \bibinfo{publisher}{ACM}, \bibinfo{address}{Montreal, QC, Canada},
  \bibinfo{pages}{41--48}.
\newblock


\bibitem[\protect\citeauthoryear{Cohn, Kruez, Matthews, Yang, Nguyen, Padilla,
  Zhou, and De~la Torre}{Cohn et~al\mbox{.}}{2009}]%
        {cohn2009detecting}
\bibfield{author}{\bibinfo{person}{Jeffrey~F. Cohn},
  \bibinfo{person}{Tomas~Simon Kruez}, \bibinfo{person}{Iain Matthews},
  \bibinfo{person}{Ying Yang}, \bibinfo{person}{Minh~Hoai Nguyen},
  \bibinfo{person}{Margara~Tejera Padilla}, \bibinfo{person}{Feng Zhou}, {and}
  \bibinfo{person}{Fernando De~la Torre}.} \bibinfo{year}{2009}\natexlab{}.
\newblock \showarticletitle{{Detecting Depression from Facial Actions and Vocal
  Prosody}}. In \bibinfo{booktitle}{\emph{Proc.\ 3rd International Conference
  on Affective Computing and Intelligent Interaction and Workshops}}.
  \bibinfo{publisher}{IEEE}, \bibinfo{address}{Amsterdam, Netherlands}.
\newblock
\newblock
\shownote{7 pages.}


\bibitem[\protect\citeauthoryear{Cordaro, Sun, Keltner, Kamble, Huddar, and
  McNeil}{Cordaro et~al\mbox{.}}{2018}]%
        {cordaro2018universals}
\bibfield{author}{\bibinfo{person}{Daniel~T. Cordaro}, \bibinfo{person}{Rui
  Sun}, \bibinfo{person}{Dacher Keltner}, \bibinfo{person}{Shanmukh Kamble},
  \bibinfo{person}{Niranjan Huddar}, {and} \bibinfo{person}{Galen McNeil}.}
  \bibinfo{year}{2018}\natexlab{}.
\newblock \showarticletitle{Universals and cultural variations in 22 emotional
  expressions across five cultures}.
\newblock \bibinfo{journal}{\emph{Emotion}}  \bibinfo{volume}{18}
  (\bibinfo{year}{2018}), \bibinfo{pages}{75--93}.
\newblock


\bibitem[\protect\citeauthoryear{Corneanu, Sim{\'o}n, Cohn, and
  Guerrero}{Corneanu et~al\mbox{.}}{2016}]%
        {Corneanu16-SOR}
\bibfield{author}{\bibinfo{person}{Ciprian~A. Corneanu},
  \bibinfo{person}{Marc~O. Sim{\'o}n}, \bibinfo{person}{Jeffrey~F. Cohn}, {and}
  \bibinfo{person}{Sergio~E. Guerrero}.} \bibinfo{year}{2016}\natexlab{}.
\newblock \showarticletitle{{Survey on RGB, 3D, Thermal, and Multimodal
  Approaches for Facial Expression Recognition: History, Trends, and
  Affect-Related Applications}}.
\newblock \bibinfo{journal}{\emph{IEEE Transactions on Pattern Analysis and
  Machine Intelligence}} \bibinfo{volume}{38}, \bibinfo{number}{8}
  (\bibinfo{date}{August} \bibinfo{year}{2016}), \bibinfo{pages}{1548--1568}.
\newblock


\bibitem[\protect\citeauthoryear{Cummins, Scherer, Krajewski, Schnieder, Epps,
  and Quatieri}{Cummins et~al\mbox{.}}{2015}]%
        {cummins2015review}
\bibfield{author}{\bibinfo{person}{Nicholas Cummins}, \bibinfo{person}{Stefan
  Scherer}, \bibinfo{person}{Jarek Krajewski}, \bibinfo{person}{Sebastian
  Schnieder}, \bibinfo{person}{Julien Epps}, {and} \bibinfo{person}{Thomas~F
  Quatieri}.} \bibinfo{year}{2015}\natexlab{}.
\newblock \showarticletitle{A review of depression and suicide risk assessment
  using speech analysis}.
\newblock \bibinfo{journal}{\emph{Speech Communication}}  \bibinfo{volume}{71}
  (\bibinfo{date}{July} \bibinfo{year}{2015}), \bibinfo{pages}{10--49}.
\newblock


\bibitem[\protect\citeauthoryear{Dautenhahn}{Dautenhahn}{2002}]%
        {dautenhahn2002origins}
\bibfield{author}{\bibinfo{person}{Kerstin Dautenhahn}.}
  \bibinfo{year}{2002}\natexlab{}.
\newblock \showarticletitle{The origins of narrative: In search of the
  transactional format of narratives in humans and other social animals}.
\newblock \bibinfo{journal}{\emph{International Journal of Cognition and
  Technology}} \bibinfo{volume}{1}, \bibinfo{number}{1} (\bibinfo{year}{2002}),
  \bibinfo{pages}{97--123}.
\newblock


\bibitem[\protect\citeauthoryear{Deng, Cummins, Schmitt, Qian, Ringeval, and
  Schuller}{Deng et~al\mbox{.}}{2017}]%
        {Deng17-SDO}
\bibfield{author}{\bibinfo{person}{Jun Deng}, \bibinfo{person}{Nicholas
  Cummins}, \bibinfo{person}{Maximilian Schmitt}, \bibinfo{person}{Kun Qian},
  \bibinfo{person}{Fabien Ringeval}, {and} \bibinfo{person}{Bj\"orn Schuller}.}
  \bibinfo{year}{2017}\natexlab{}.
\newblock \showarticletitle{{Speech-based diagnosis of autism spectrum
  condition by generative adversarial network representations}}. In
  \bibinfo{booktitle}{\emph{{Proc.\, 7th International Conference on Digital
  Health (DH)}}}. \bibinfo{publisher}{ACM}, \bibinfo{address}{London, UK},
  \bibinfo{pages}{53--57}.
\newblock


\bibitem[\protect\citeauthoryear{DeVault, Artstein, Benn, Dey, Fast, Gainer,
  Georgila, Gratch, Hartholt, Lhommet, Lucas, Marsella, Morbini, Nazarian,
  Scherer, Stratou, Suri, Traum, Wood, Xu, Rizzo, and Morency}{DeVault
  et~al\mbox{.}}{2014}]%
        {DeVaultEtAl2014_SVH}
\bibfield{author}{\bibinfo{person}{David DeVault}, \bibinfo{person}{Ron
  Artstein}, \bibinfo{person}{Grace Benn}, \bibinfo{person}{Teresa Dey},
  \bibinfo{person}{Ed Fast}, \bibinfo{person}{Alesia Gainer},
  \bibinfo{person}{Kallirroi Georgila}, \bibinfo{person}{Jonathan Gratch},
  \bibinfo{person}{Arno Hartholt}, \bibinfo{person}{Margaux Lhommet},
  \bibinfo{person}{Gale Lucas}, \bibinfo{person}{Stacy Marsella},
  \bibinfo{person}{Fabrizio Morbini}, \bibinfo{person}{Angela Nazarian},
  \bibinfo{person}{Stefan Scherer}, \bibinfo{person}{Giota Stratou},
  \bibinfo{person}{Apar Suri}, \bibinfo{person}{David Traum},
  \bibinfo{person}{Rachel Wood}, \bibinfo{person}{Yuyu Xu},
  \bibinfo{person}{Alberto Rizzo}, {and} \bibinfo{person}{Louis-Philippe
  Morency}.} \bibinfo{year}{2014}\natexlab{}.
\newblock \showarticletitle{{SimSensei Kiosk: A Virtual Human Interviewer for
  Healthcare Decision Support}}. In \bibinfo{booktitle}{\emph{Proc.\
  International Conference on Autonomous Agents and Multi-Agent Systems, AAMAS
  2014}}. \bibinfo{publisher}{ACM}, \bibinfo{address}{Paris, France},
  \bibinfo{pages}{1061--1068}.
\newblock


\bibitem[\protect\citeauthoryear{D'{M}ello and Kory}{D'{M}ello and
  Kory}{2015}]%
        {Dmello15-ARA}
\bibfield{author}{\bibinfo{person}{Sidney~K. D'{M}ello} {and}
  \bibinfo{person}{Jacqueline Kory}.} \bibinfo{year}{2015}\natexlab{}.
\newblock \showarticletitle{{A Review and Meta-Analysis of Multimodal Affect
  Detection Systems}}.
\newblock \bibinfo{journal}{\emph{Comput. Surveys}} \bibinfo{volume}{47},
  \bibinfo{number}{3} (\bibinfo{date}{February} \bibinfo{year}{2015}).
\newblock
\newblock
\shownote{Article 43, 36 pages.}


\bibitem[\protect\citeauthoryear{Ekman}{Ekman}{1971}]%
        {Ekamn71-UAC}
\bibfield{author}{\bibinfo{person}{Paul Ekman}.}
  \bibinfo{year}{1971}\natexlab{}.
\newblock \showarticletitle{Universals and cultural differences in facial
  expressions of emotion}. In \bibinfo{booktitle}{\emph{{Nebraska Symposium on
  Motivation}}}, Vol.~\bibinfo{volume}{19}. \bibinfo{publisher}{University of
  Nebraska Press}, \bibinfo{address}{Lincoln, NE}, \bibinfo{pages}{207--283}.
\newblock


\bibitem[\protect\citeauthoryear{Elfenbein and Ambady}{Elfenbein and
  Ambady}{2002}]%
        {Elfenbein02-otu}
\bibfield{author}{\bibinfo{person}{Hillary~Anger Elfenbein} {and}
  \bibinfo{person}{Nalini Ambady}.} \bibinfo{year}{2002}\natexlab{}.
\newblock \showarticletitle{On the universality and cultural specificity of
  emotion recognition: A meta-analysis}.
\newblock \bibinfo{journal}{\emph{Psychological Bulletin}}
  \bibinfo{volume}{128}, \bibinfo{number}{2} (\bibinfo{year}{2002}),
  \bibinfo{pages}{203--235}.
\newblock


\bibitem[\protect\citeauthoryear{Esposito, Esposito, and Vogel}{Esposito
  et~al\mbox{.}}{2015}]%
        {Esposito15-NAC}
\bibfield{author}{\bibinfo{person}{Anna Esposito},
  \bibinfo{person}{Antonietta~M. Esposito}, {and} \bibinfo{person}{Carl
  Vogel}.} \bibinfo{year}{2015}\natexlab{}.
\newblock \showarticletitle{Needs and challenges in human computer interaction
  for processing social emotional information}.
\newblock \bibinfo{journal}{\emph{Pattern Recognition Letters}}
  \bibinfo{volume}{66} (\bibinfo{date}{November} \bibinfo{year}{2015}),
  \bibinfo{pages}{41--51}.
\newblock
Issue C.


\bibitem[\protect\citeauthoryear{Eyben, Scherer, Schuller, Sundberg, Andr{\'e},
  Busso, Devillers, Epps, Laukka, Narayanan, and Truong}{Eyben
  et~al\mbox{.}}{2016}]%
        {Eyben16-TGM}
\bibfield{author}{\bibinfo{person}{Florian Eyben}, \bibinfo{person}{Klaus~R.
  Scherer}, \bibinfo{person}{Bj{\"o}rn Schuller}, \bibinfo{person}{Johan
  Sundberg}, \bibinfo{person}{Elisabeth Andr{\'e}}, \bibinfo{person}{Carlos
  Busso}, \bibinfo{person}{Laurence Devillers}, \bibinfo{person}{Julien Epps},
  \bibinfo{person}{Petri Laukka}, \bibinfo{person}{Shrikanth~S. Narayanan},
  {and} \bibinfo{person}{Khiet~P. Truong}.} \bibinfo{year}{2016}\natexlab{}.
\newblock \showarticletitle{{The Geneva Minimalistic Acoustic Parameter Set
  (GeMAPS) for Voice Research and Affective Computing}}.
\newblock \bibinfo{journal}{\emph{IEEE Transactions on Affective Computing}}
  \bibinfo{volume}{7}, \bibinfo{number}{2} (\bibinfo{date}{April}
  \bibinfo{year}{2016}), \bibinfo{pages}{190--202}.
\newblock


\bibitem[\protect\citeauthoryear{Eyben, Weninger, Gro{\ss}, and Schuller}{Eyben
  et~al\mbox{.}}{2013}]%
        {Eyben13-RDI}
\bibfield{author}{\bibinfo{person}{Florian Eyben}, \bibinfo{person}{Felix
  Weninger}, \bibinfo{person}{Florian Gro{\ss}}, {and} \bibinfo{person}{Bj\"orn
  Schuller}.} \bibinfo{year}{2013}\natexlab{}.
\newblock \showarticletitle{{Recent Developments in openSMILE, the Munich
  Open-Source Multimedia Feature Extractor}}. In
  \bibinfo{booktitle}{\emph{{Proc.\ 21st ACM International Conference on
  Multimedia (ACM MM)}}}. \bibinfo{publisher}{ACM},
  \bibinfo{address}{Barcelona, Spain}, \bibinfo{pages}{835--838}.
\newblock


\bibitem[\protect\citeauthoryear{Feraru, Schuller, and Schuller}{Feraru
  et~al\mbox{.}}{2015}]%
        {Feraru15-CAE}
\bibfield{author}{\bibinfo{person}{{Silvia Monica} Feraru},
  \bibinfo{person}{Dagmar Schuller}, {and} \bibinfo{person}{Bj\"orn Schuller}.}
  \bibinfo{year}{2015}\natexlab{}.
\newblock \showarticletitle{{Cross-Language Acoustic Emotion Recognition: An
  Overview and Some Tendencies}}. In \bibinfo{booktitle}{\emph{{Proc.\ 6th
  Biannual Conference on Affective Computing and Intelligent Interaction
  (ACII)}}}. \bibinfo{publisher}{IEEE}, \bibinfo{address}{Xi'an, P.\,R.\
  China}, \bibinfo{pages}{125--131}.
\newblock


\bibitem[\protect\citeauthoryear{Gong and Poellabauer}{Gong and
  Poellabauer}{2017}]%
        {Gong17-TMB}
\bibfield{author}{\bibinfo{person}{Yuan Gong} {and} \bibinfo{person}{Christian
  Poellabauer}.} \bibinfo{year}{2017}\natexlab{}.
\newblock \showarticletitle{{Topic Modeling Based Multi-modal Depression
  Detection}}. In \bibinfo{booktitle}{\emph{{Proc.\ 7th International Workshop
  on Audio/Visual Emotion Challenge (AVEC)}}}. \bibinfo{publisher}{ACM},
  \bibinfo{address}{Mountain View (CA), USA}, \bibinfo{pages}{69--76}.
\newblock


\bibitem[\protect\citeauthoryear{Gratch, Artstein, Lucas, Stratou, Scherer,
  Nazarian, Wood, Boberg, DeVault, Marsella, Traum, Rizzo, and Morency}{Gratch
  et~al\mbox{.}}{2014}]%
        {gratchetal14}
\bibfield{author}{\bibinfo{person}{Jonathan Gratch}, \bibinfo{person}{Ron
  Artstein}, \bibinfo{person}{Gale Lucas}, \bibinfo{person}{Giota Stratou},
  \bibinfo{person}{Stefan Scherer}, \bibinfo{person}{Angela Nazarian},
  \bibinfo{person}{Rachel Wood}, \bibinfo{person}{Jill Boberg},
  \bibinfo{person}{David DeVault}, \bibinfo{person}{Stacy Marsella},
  \bibinfo{person}{David Traum}, \bibinfo{person}{Skip Rizzo}, {and}
  \bibinfo{person}{Louis-Philippe Morency}.} \bibinfo{year}{2014}\natexlab{}.
\newblock \showarticletitle{{The Distress Analysis Interview Corpus of human
  and computer interviews}}. In \bibinfo{booktitle}{\emph{Proc.\ 9th
  International Conference on Language Resources and Evaluation, LREC 2014}}.
  \bibinfo{publisher}{ELRA}, \bibinfo{address}{Reykjavik, Iceland},
  \bibinfo{pages}{3123--3128}.
\newblock


\bibitem[\protect\citeauthoryear{He, Zhang, Ren, and Sun}{He
  et~al\mbox{.}}{2016}]%
        {he2016deep}
\bibfield{author}{\bibinfo{person}{Kaiming He}, \bibinfo{person}{Xiangyu
  Zhang}, \bibinfo{person}{Shaoqing Ren}, {and} \bibinfo{person}{Jian Sun}.}
  \bibinfo{year}{2016}\natexlab{}.
\newblock \showarticletitle{{Deep Residual Learning for Image Recognition}}. In
  \bibinfo{booktitle}{\emph{Proc.\ IEEE Conference on Computer Vision and
  Pattern Recognition}}. \bibinfo{publisher}{IEEE}, \bibinfo{address}{Las
  Vegas, NV}, \bibinfo{pages}{770--778}.
\newblock


\bibitem[\protect\citeauthoryear{Houben, Noortgate, and Kuppens}{Houben
  et~al\mbox{.}}{2015}]%
        {Houben15-TRB}
\bibfield{author}{\bibinfo{person}{Marlies Houben}, \bibinfo{person}{Wim
  Van~Den Noortgate}, {and} \bibinfo{person}{Peter Kuppens}.}
  \bibinfo{year}{2015}\natexlab{}.
\newblock \showarticletitle{The relation between short term emotion dynamics
  and psychological well-being: A meta-analysis}.
\newblock \bibinfo{journal}{\emph{Psychological Bulletin}}
  \bibinfo{volume}{141}, \bibinfo{number}{4} (\bibinfo{date}{July}
  \bibinfo{year}{2015}), \bibinfo{pages}{901--930}.
\newblock


\bibitem[\protect\citeauthoryear{Huang, Liu, van~der Maaten, and
  Weinberger}{Huang et~al\mbox{.}}{2017}]%
        {Huang_2017_CVPR}
\bibfield{author}{\bibinfo{person}{Gao Huang}, \bibinfo{person}{Zhuang Liu},
  \bibinfo{person}{Laurens van~der Maaten}, {and} \bibinfo{person}{Kilian~Q.
  Weinberger}.} \bibinfo{year}{2017}\natexlab{}.
\newblock \showarticletitle{Densely Connected Convolutional Networks}. In
  \bibinfo{booktitle}{\emph{The IEEE Conference on Computer Vision and Pattern
  Recognition (CVPR)}}. \bibinfo{publisher}{IEEE}, \bibinfo{address}{Honolulu,
  HW}, \bibinfo{pages}{4700--4708}.
\newblock


\bibitem[\protect\citeauthoryear{Huang, Li, Tao, Lian, Niu, and Yang}{Huang
  et~al\mbox{.}}{2018}]%
        {Huang18-MCE}
\bibfield{author}{\bibinfo{person}{Jian Huang}, \bibinfo{person}{Ya Li},
  \bibinfo{person}{Jianhua Tao}, \bibinfo{person}{Zheng Lian},
  \bibinfo{person}{Mingyue Niu}, {and} \bibinfo{person}{Minghao Yang}.}
  \bibinfo{year}{2018}\natexlab{}.
\newblock \showarticletitle{Multimodal Continuous Emotion Recognition with Data
  Augmentation Using Recurrent Neural Networks}. In
  \bibinfo{booktitle}{\emph{{Proc.\ 8th International Workshop on Audio/Visual
  Emotion Challenge, AVEC'18}}}. \bibinfo{publisher}{ACM},
  \bibinfo{address}{Seoul, South Korea}, \bibinfo{pages}{57--64}.
\newblock


\bibitem[\protect\citeauthoryear{Joshi, Goecke, Alghowinem, Dhall, Wagner,
  Epps, Parker, and Breakspear}{Joshi et~al\mbox{.}}{2013}]%
        {joshi2013multimodal}
\bibfield{author}{\bibinfo{person}{Jyoti Joshi}, \bibinfo{person}{Roland
  Goecke}, \bibinfo{person}{Sharifa Alghowinem}, \bibinfo{person}{Abhinav
  Dhall}, \bibinfo{person}{Michael Wagner}, \bibinfo{person}{Julien Epps},
  \bibinfo{person}{Gordon Parker}, {and} \bibinfo{person}{Michael Breakspear}.}
  \bibinfo{year}{2013}\natexlab{}.
\newblock \showarticletitle{Multimodal assistive technologies for depression
  diagnosis and monitoring}.
\newblock \bibinfo{journal}{\emph{Journal on Multimodal User Interfaces}}
  \bibinfo{volume}{7}, \bibinfo{number}{3} (\bibinfo{year}{2013}),
  \bibinfo{pages}{217--228}.
\newblock


\bibitem[\protect\citeauthoryear{Kaya and Karpov}{Kaya and Karpov}{2018}]%
        {Kaya18-EAE}
\bibfield{author}{\bibinfo{person}{Heysem Kaya} {and}
  \bibinfo{person}{Alexey~A. Karpov}.} \bibinfo{year}{2018}\natexlab{}.
\newblock \showarticletitle{Efficient and effective strategies for cross-corpus
  acoustic emotion recognition}.
\newblock \bibinfo{journal}{\emph{Neurocomputing}}  \bibinfo{volume}{275}
  (\bibinfo{date}{January} \bibinfo{year}{2018}), \bibinfo{pages}{1028--034}.
\newblock


\bibitem[\protect\citeauthoryear{Kollias, Tzirakis, Nicolaou, Papaioannou,
  Zhao, Schuller, Kotsia, and Zafeiriou}{Kollias et~al\mbox{.}}{2019}]%
        {kollias2019deep}
\bibfield{author}{\bibinfo{person}{Dimitrios Kollias},
  \bibinfo{person}{Panagiotis Tzirakis}, \bibinfo{person}{Mihalis~A. Nicolaou},
  \bibinfo{person}{Athanasios Papaioannou}, \bibinfo{person}{Guoying Zhao},
  \bibinfo{person}{Bj{\"o}rn Schuller}, \bibinfo{person}{Irene Kotsia}, {and}
  \bibinfo{person}{Stefanos Zafeiriou}.} \bibinfo{year}{2019}\natexlab{}.
\newblock \showarticletitle{Deep affect prediction in-the-wild: Aff-wild
  database and challenge, deep architectures, and beyond}.
\newblock \bibinfo{journal}{\emph{International Journal of Computer Vision}}
  \bibinfo{volume}{127}, \bibinfo{number}{6} (\bibinfo{year}{2019}),
  \bibinfo{pages}{907--929}.
\newblock


\bibitem[\protect\citeauthoryear{Kossaifi, Walecki, Panagakis, Shen, Schmitt,
  Ringeval, Han, Pandit, Schuller, Star, Hajiyev, and Pantic}{Kossaifi
  et~al\mbox{.}}{2019}]%
        {kossaifi19-sda}
\bibfield{author}{\bibinfo{person}{Jean Kossaifi}, \bibinfo{person}{Robert
  Walecki}, \bibinfo{person}{Yannis Panagakis}, \bibinfo{person}{Jie Shen},
  \bibinfo{person}{Maximilian Schmitt}, \bibinfo{person}{Fabien Ringeval},
  \bibinfo{person}{Jing Han}, \bibinfo{person}{Vedhas Pandit},
  \bibinfo{person}{Bjorn Schuller}, \bibinfo{person}{Kam Star},
  \bibinfo{person}{Elnar Hajiyev}, {and} \bibinfo{person}{Maja Pantic}.}
  \bibinfo{year}{2019}\natexlab{}.
\newblock \bibinfo{title}{{SEWA DB: A Rich Database for Audio-Visual Emotion
  and Sentiment Research in the Wild}}.
\newblock \bibinfo{howpublished}{\url{https://arxiv.org/abs/1901.02839}}.
\newblock
\newblock
\shownote{17 pages.}


\bibitem[\protect\citeauthoryear{Koval, Kuppens, Allen, and Sheeber}{Koval
  et~al\mbox{.}}{2012}]%
        {koval2012getting}
\bibfield{author}{\bibinfo{person}{Peter Koval}, \bibinfo{person}{Peter
  Kuppens}, \bibinfo{person}{Nicholas~B. Allen}, {and} \bibinfo{person}{Lisa
  Sheeber}.} \bibinfo{year}{2012}\natexlab{}.
\newblock \showarticletitle{Getting stuck in depression: The roles of
  rumination and emotional inertia}.
\newblock \bibinfo{journal}{\emph{Cognition \& Emotion}} \bibinfo{volume}{26},
  \bibinfo{number}{8} (\bibinfo{year}{2012}), \bibinfo{pages}{1412--1427}.
\newblock


\bibitem[\protect\citeauthoryear{Koval, Pe, Meers, and Kuppens}{Koval
  et~al\mbox{.}}{2013}]%
        {koval2013affect}
\bibfield{author}{\bibinfo{person}{Peter Koval}, \bibinfo{person}{Madeline~L.
  Pe}, \bibinfo{person}{Kristof Meers}, {and} \bibinfo{person}{Peter Kuppens}.}
  \bibinfo{year}{2013}\natexlab{}.
\newblock \showarticletitle{Affect dynamics in relation to depressive symptoms:
  Variable, unstable or inert?}
\newblock \bibinfo{journal}{\emph{Emotion}} \bibinfo{volume}{13},
  \bibinfo{number}{6} (\bibinfo{year}{2013}), \bibinfo{pages}{1132}.
\newblock


\bibitem[\protect\citeauthoryear{Krizhevsky, Sutskever, and Hinton}{Krizhevsky
  et~al\mbox{.}}{2012}]%
        {Krizhevsky2012}
\bibfield{author}{\bibinfo{person}{Alex Krizhevsky}, \bibinfo{person}{Ilya
  Sutskever}, {and} \bibinfo{person}{Geoffrey~E. Hinton}.}
  \bibinfo{year}{2012}\natexlab{}.
\newblock \showarticletitle{{ImageNet Classification with Deep Convolutional
  Neural Networks}}.
\newblock In \bibinfo{booktitle}{\emph{{Advances in Neural Information
  Processing Systems 25}}}, \bibfield{editor}{\bibinfo{person}{F.~Pereira},
  \bibinfo{person}{C.~J.~C. Burges}, \bibinfo{person}{L.~Bottou}, {and}
  \bibinfo{person}{K.~Q. Weinberger}} (Eds.). \bibinfo{publisher}{Curran
  Associates, Inc.}, \bibinfo{address}{Lake Tahoe, NV},
  \bibinfo{pages}{1097--1105}.
\newblock


\bibitem[\protect\citeauthoryear{Kuppens, Allen, and Sheeber}{Kuppens
  et~al\mbox{.}}{2010}]%
        {kuppens2010emotional}
\bibfield{author}{\bibinfo{person}{Peter Kuppens}, \bibinfo{person}{Nicholas~B.
  Allen}, {and} \bibinfo{person}{Lisa~B. Sheeber}.}
  \bibinfo{year}{2010}\natexlab{}.
\newblock \showarticletitle{Emotional inertia and psychological maladjustment}.
\newblock \bibinfo{journal}{\emph{Psychological Science}} \bibinfo{volume}{21},
  \bibinfo{number}{7} (\bibinfo{year}{2010}), \bibinfo{pages}{984--991}.
\newblock


\bibitem[\protect\citeauthoryear{Li}{Li}{1989}]%
        {Li89-ACC}
\bibfield{author}{\bibinfo{person}{Lin Li}.} \bibinfo{year}{1989}\natexlab{}.
\newblock \showarticletitle{A concordance correlation coefficient to evaluate
  reproducibility}.
\newblock \bibinfo{journal}{\emph{Biometrics}} \bibinfo{volume}{45},
  \bibinfo{number}{1} (\bibinfo{date}{March} \bibinfo{year}{1989}),
  \bibinfo{pages}{255--268}.
\newblock


\bibitem[\protect\citeauthoryear{Lotfian and Busso}{Lotfian and Busso}{2019}]%
        {Lotfian19-CLF}
\bibfield{author}{\bibinfo{person}{Reza Lotfian} {and} \bibinfo{person}{Carlos
  Busso}.} \bibinfo{year}{2019}\natexlab{}.
\newblock \showarticletitle{Curriculum Learning for Speech Emotion Recognition
  from Crowdsourced Labels}.
\newblock \bibinfo{journal}{\emph{IEEE Transactions on Audio, Speech \&
  Language Processing}} \bibinfo{volume}{27}, \bibinfo{number}{4}
  (\bibinfo{year}{2019}), \bibinfo{pages}{815--826}.
\newblock


\bibitem[\protect\citeauthoryear{Maturana and Varela}{Maturana and
  Varela}{1987}]%
        {maturana1987tree}
\bibfield{author}{\bibinfo{person}{Humberto~R. Maturana} {and}
  \bibinfo{person}{Francisco~J. Varela}.} \bibinfo{year}{1987}\natexlab{}.
\newblock \bibinfo{booktitle}{\emph{{Tree of Knowledge: The Biological Roots of
  Human Understanding}}}.
\newblock \bibinfo{publisher}{New Science Library/Shambhala Publications},
  \bibinfo{address}{Boston, MA}.
\newblock


\bibitem[\protect\citeauthoryear{Morales, Scherer, and Levitan}{Morales
  et~al\mbox{.}}{2017}]%
        {morales2017cross}
\bibfield{author}{\bibinfo{person}{Michelle Morales}, \bibinfo{person}{Stefan
  Scherer}, {and} \bibinfo{person}{Rivka Levitan}.}
  \bibinfo{year}{2017}\natexlab{}.
\newblock \showarticletitle{{A Cross-modal Review of Indicators for Depression
  Detection Systems}}. In \bibinfo{booktitle}{\emph{Proc\, 4th Workshop on
  Computational Linguistics and Clinical Psychology -- From Linguistic Signal
  to Clinical Reality}}. \bibinfo{publisher}{ACL}, \bibinfo{address}{Vancouver,
  BC}, \bibinfo{pages}{1--12}.
\newblock


\bibitem[\protect\citeauthoryear{Organization}{Organization}{2017}]%
        {world2017depression}
\bibfield{author}{\bibinfo{person}{World~Health Organization}.}
  \bibinfo{year}{2017}\natexlab{}.
\newblock \bibinfo{booktitle}{\emph{{Depression and Other Common Mental
  Disorders: Global Health Estimates}}}.
\newblock \bibinfo{type}{{T}echnical {R}eport}. \bibinfo{institution}{World
  Health Organization}.
\newblock
\newblock
\shownote{Licence: CC BY-NC-SA 3.0 IGO.}


\bibitem[\protect\citeauthoryear{Pandit and Schuller}{Pandit and
  Schuller}{2019}]%
        {Pandit19-OMM}
\bibfield{author}{\bibinfo{person}{Vedhas Pandit} {and}
  \bibinfo{person}{Bj\"orn Schuller}.} \bibinfo{year}{2019}\natexlab{}.
\newblock \bibinfo{title}{{On Many-to-Many Mapping Between Concordance
  Correlation Coefficient and Mean Square Error}}.
\newblock \bibinfo{howpublished}{\url{https://arxiv.org/abs/1902.05180}}.
\newblock
\newblock
\shownote{23 pages.}


\bibitem[\protect\citeauthoryear{Pantic, Sebe, Cohn, and Huang}{Pantic
  et~al\mbox{.}}{2005}]%
        {Pantic05-AMH}
\bibfield{author}{\bibinfo{person}{Maja Pantic}, \bibinfo{person}{Nicu Sebe},
  \bibinfo{person}{Jeffrey~F. Cohn}, {and} \bibinfo{person}{Thomas Huang}.}
  \bibinfo{year}{2005}\natexlab{}.
\newblock \showarticletitle{Affective Multimodal Human-computer Interaction}.
  In \bibinfo{booktitle}{\emph{Proc.\ 13th Annual ACM International Conference
  on Multimedia}}. \bibinfo{publisher}{ACM}, \bibinfo{address}{Singapore,
  Singapore}, \bibinfo{pages}{669--676}.
\newblock


\bibitem[\protect\citeauthoryear{Pennebaker, Mehl, and Niederhoffer}{Pennebaker
  et~al\mbox{.}}{2003}]%
        {pennebaker2003psychological}
\bibfield{author}{\bibinfo{person}{James~W Pennebaker},
  \bibinfo{person}{Matthias~R Mehl}, {and} \bibinfo{person}{Kate~G
  Niederhoffer}.} \bibinfo{year}{2003}\natexlab{}.
\newblock \showarticletitle{{Psychological Aspects of Natural Language Use: Our
  Words, Our Selves}}.
\newblock \bibinfo{journal}{\emph{Annual Review of Psychology}}
  \bibinfo{volume}{54}, \bibinfo{number}{1} (\bibinfo{year}{2003}),
  \bibinfo{pages}{547--577}.
\newblock


\bibitem[\protect\citeauthoryear{Ramirez-Esparza, Chung, Kacewicz, and
  Pennebaker}{Ramirez-Esparza et~al\mbox{.}}{2008}]%
        {ramirez2008psychology}
\bibfield{author}{\bibinfo{person}{Nairan Ramirez-Esparza},
  \bibinfo{person}{Cindy~K Chung}, \bibinfo{person}{Ewa Kacewicz}, {and}
  \bibinfo{person}{James~W Pennebaker}.} \bibinfo{year}{2008}\natexlab{}.
\newblock \showarticletitle{The Psychology of Word Use in Depression Forums in
  English and in Spanish: Texting Two Text Analytic Approaches}. In
  \bibinfo{booktitle}{\emph{International Conference on Weblogs and Social
  Media}}. \bibinfo{publisher}{AAAI}, \bibinfo{address}{Seattle, WA},
  \bibinfo{pages}{102--108}.
\newblock


\bibitem[\protect\citeauthoryear{Rathner, Djamali, Terhorst, Schuller, Cummins,
  Salamon, Hunger-Schoppe, and Baumeister}{Rathner et~al\mbox{.}}{2018a}]%
        {rathner2018did}
\bibfield{author}{\bibinfo{person}{Eva-Maria Rathner}, \bibinfo{person}{Julia
  Djamali}, \bibinfo{person}{Yannik Terhorst}, \bibinfo{person}{Bj{\"o}rn
  Schuller}, \bibinfo{person}{Nicholas Cummins}, \bibinfo{person}{Gudrun
  Salamon}, \bibinfo{person}{Christina Hunger-Schoppe}, {and}
  \bibinfo{person}{Harald Baumeister}.} \bibinfo{year}{2018}\natexlab{a}.
\newblock \showarticletitle{How Did You like 2017? Detection of Language
  Markers of Depression and Narcissism in Personal Narratives}. In
  \bibinfo{booktitle}{\emph{Proc.\ of INTERSPEECH 2018, 19th Annual Conference
  of the International Speech Communication Association}}.
  \bibinfo{publisher}{ISCA}, \bibinfo{address}{Hyderabad, India},
  \bibinfo{pages}{3388--3392}.
\newblock


\bibitem[\protect\citeauthoryear{Rathner, Terhorst, Cummins, Schuller, and
  Baumeister}{Rathner et~al\mbox{.}}{2018b}]%
        {rathner18som}
\bibfield{author}{\bibinfo{person}{Eva-Maria Rathner}, \bibinfo{person}{Yannik
  Terhorst}, \bibinfo{person}{Nicholas Cummins}, \bibinfo{person}{Bj\"orn
  Schuller}, {and} \bibinfo{person}{Harald Baumeister}.}
  \bibinfo{year}{2018}\natexlab{b}.
\newblock \showarticletitle{State of Mind: Classification through Self-reported
  Affect and Word Use in Speech}. In \bibinfo{booktitle}{\emph{Proc.\ of
  INTERSPEECH 2018, 19th Annual Conference of the International Speech
  Communication Association}}. \bibinfo{publisher}{ISCA},
  \bibinfo{address}{Hyderabad, India}, \bibinfo{pages}{267--271}.
\newblock


\bibitem[\protect\citeauthoryear{Ringeval, Schuller, ddy Cowie, Kaya, Schmitt,
  Amiriparian, Cummins, Lalanne, Michaud, Ciftci, G\"ulec, Salah, and
  Pantic}{Ringeval et~al\mbox{.}}{2018a}]%
        {Ringeval18-A2W}
\bibfield{author}{\bibinfo{person}{Fabien Ringeval}, \bibinfo{person}{Bj\"orn
  Schuller}, \bibinfo{person}{Michel Valstarand~Ro ddy Cowie},
  \bibinfo{person}{Heysem Kaya}, \bibinfo{person}{Maximilian Schmitt},
  \bibinfo{person}{Shahin Amiriparian}, \bibinfo{person}{Nicholas Cummins},
  \bibinfo{person}{Dennis Lalanne}, \bibinfo{person}{Adrien Michaud},
  \bibinfo{person}{Elvan Ciftci}, \bibinfo{person}{H\"useyin G\"ulec},
  \bibinfo{person}{Albert~Ali Salah}, {and} \bibinfo{person}{Maja Pantic}.}
  \bibinfo{year}{2018}\natexlab{a}.
\newblock \showarticletitle{{AVEC 2018 Workshop and Challenge: Bipolar Disorder
  and Cross-Cultural Affect Recognition}}. In \bibinfo{booktitle}{\emph{{Proc.\
  8th International Workshop on Audio/Visual Emotion Challenge, AVEC'18}}}.
  \bibinfo{publisher}{ACM}, \bibinfo{address}{Seoul, South Korea},
  \bibinfo{pages}{3--13}.
\newblock


\bibitem[\protect\citeauthoryear{Ringeval, Schuller, Valstar, Cowie, and
  Pantic}{Ringeval et~al\mbox{.}}{2015}]%
        {Ringeval15-A2Ta}
\bibfield{author}{\bibinfo{person}{Fabien Ringeval}, \bibinfo{person}{Bj\"orn
  Schuller}, \bibinfo{person}{Michel Valstar}, \bibinfo{person}{Roddy Cowie},
  {and} \bibinfo{person}{Maja Pantic}.} \bibinfo{year}{2015}\natexlab{}.
\newblock \showarticletitle{{AVEC 2015 -- The 5th International Audio/Visual
  Emotion Challenge and Workshop}}. In \bibinfo{booktitle}{\emph{{Proc.\ 23rd
  ACM International Conference on Multimedia, MM 2015}}}.
  \bibinfo{publisher}{ACM}, \bibinfo{address}{Brisbane, Australia},
  \bibinfo{pages}{1335--1336}.
\newblock


\bibitem[\protect\citeauthoryear{Ringeval, Schuller, Valstar, Cowie, and
  Pantic}{Ringeval et~al\mbox{.}}{2017a}]%
        {Ringeval17-SFA}
\bibfield{author}{\bibinfo{person}{Fabien Ringeval}, \bibinfo{person}{Bj\"orn
  Schuller}, \bibinfo{person}{Michel Valstar}, \bibinfo{person}{Roddy Cowie},
  {and} \bibinfo{person}{Maja Pantic}.} \bibinfo{year}{2017}\natexlab{a}.
\newblock \showarticletitle{{Summary for AVEC 2017 -- Real-life Depression, and
  Affect Recognition Challenge sand Workshop}}. In
  \bibinfo{booktitle}{\emph{{Proc.\ 25th ACM International Conference on
  Multimedia (ACM MM)}}}. \bibinfo{publisher}{ACM}, \bibinfo{address}{Mountain
  View, CA, USA}, \bibinfo{pages}{1963--1964}.
\newblock


\bibitem[\protect\citeauthoryear{Ringeval, Schuller, Valstar, Cowie, and
  Pantic}{Ringeval et~al\mbox{.}}{2018b}]%
        {Ringeval18-SFA}
\bibfield{author}{\bibinfo{person}{Fabien Ringeval}, \bibinfo{person}{Bj\"orn
  Schuller}, \bibinfo{person}{Michel Valstar}, \bibinfo{person}{Roddy Cowie},
  {and} \bibinfo{person}{Maja Pantic}.} \bibinfo{year}{2018}\natexlab{b}.
\newblock \showarticletitle{{Summary for AVEC 2018: Bipolar Disorder and
  Cross-Cultural Affect Recognition}}. In \bibinfo{booktitle}{\emph{{Proc.\
  26th ACM International Conference on Multimedia, MM 2018}}}.
  \bibinfo{publisher}{ACM}, \bibinfo{address}{Seoul, South Korea},
  \bibinfo{pages}{2111--2112}.
\newblock


\bibitem[\protect\citeauthoryear{Ringeval, Schuller, Valstar, Gratch, Cowie,
  Scherer, Mozgai, Cummins, and Pantic}{Ringeval et~al\mbox{.}}{2017b}]%
        {ringeval17a2r}
\bibfield{author}{\bibinfo{person}{Fabien Ringeval}, \bibinfo{person}{Bj\"orn
  Schuller}, \bibinfo{person}{Michel Valstar}, \bibinfo{person}{Jonathan
  Gratch}, \bibinfo{person}{Roddy Cowie}, \bibinfo{person}{Stefan Scherer},
  \bibinfo{person}{Sharon Mozgai}, \bibinfo{person}{Nicholas Cummins}, {and}
  \bibinfo{person}{Maja Pantic}.} \bibinfo{year}{2017}\natexlab{b}.
\newblock \showarticletitle{{AVEC 2017 -- Real-life Depression, and Affect
  Recognition Workshop and Challenge}}. In \bibinfo{booktitle}{\emph{{Proc.\
  7th International Workshop on Audio/Visual Emotion Challenge (AVEC)}}}.
  \bibinfo{publisher}{ACM}, \bibinfo{address}{Mountain View, CA, USA},
  \bibinfo{pages}{3--9}.
\newblock


\bibitem[\protect\citeauthoryear{Russell}{Russell}{2003}]%
        {russell2003core}
\bibfield{author}{\bibinfo{person}{James~A Russell}.}
  \bibinfo{year}{2003}\natexlab{}.
\newblock \showarticletitle{Core affect and the psychological construction of
  emotion.}
\newblock \bibinfo{journal}{\emph{Psychological review}} \bibinfo{volume}{110},
  \bibinfo{number}{1} (\bibinfo{year}{2003}), \bibinfo{pages}{145}.
\newblock


\bibitem[\protect\citeauthoryear{Russell and Lanius}{Russell and
  Lanius}{1984}]%
        {russell1984adaptation}
\bibfield{author}{\bibinfo{person}{James~A Russell} {and}
  \bibinfo{person}{Ulrich~F Lanius}.} \bibinfo{year}{1984}\natexlab{}.
\newblock \showarticletitle{Adaptation level and the affective appraisal of
  environments}.
\newblock \bibinfo{journal}{\emph{Journal of Environmental Psychology}}
  \bibinfo{volume}{4}, \bibinfo{number}{2} (\bibinfo{year}{1984}),
  \bibinfo{pages}{119--135}.
\newblock


\bibitem[\protect\citeauthoryear{Sagha, Deng, Gavryukova, Han, and
  Schuller}{Sagha et~al\mbox{.}}{2016}]%
        {Sagha16-CLS}
\bibfield{author}{\bibinfo{person}{Hesam Sagha}, \bibinfo{person}{Jun Deng},
  \bibinfo{person}{Maryna Gavryukova}, \bibinfo{person}{Jing Han}, {and}
  \bibinfo{person}{Bj\"orn Schuller}.} \bibinfo{year}{2016}\natexlab{}.
\newblock \showarticletitle{{Cross lingual speech emotion recognition using
  canonical correlation analysis on principal component subspace}}. In
  \bibinfo{booktitle}{\emph{{Proc.\ 41st IEEE International Conference on
  Acoustics, Speech and Signal Processing (ICASSP) }}}.
  \bibinfo{publisher}{IEEE}, \bibinfo{address}{Shanghai, P.\,R.\ China},
  \bibinfo{pages}{5800--5804}.
\newblock


\bibitem[\protect\citeauthoryear{Sapolsky}{Sapolsky}{2004}]%
        {sapolsky2004social}
\bibfield{author}{\bibinfo{person}{Robert~M Sapolsky}.}
  \bibinfo{year}{2004}\natexlab{}.
\newblock \showarticletitle{Social status and health in humans and other
  animals}.
\newblock \bibinfo{journal}{\emph{Annu. Rev. Anthropol.}}  \bibinfo{volume}{33}
  (\bibinfo{year}{2004}), \bibinfo{pages}{393--418}.
\newblock


\bibitem[\protect\citeauthoryear{Scherer, Banse, and Wallbott}{Scherer
  et~al\mbox{.}}{2001}]%
        {Scherer01-EIF}
\bibfield{author}{\bibinfo{person}{Klaus~R. Scherer}, \bibinfo{person}{Rainer
  Banse}, {and} \bibinfo{person}{Harald~G. Wallbott}.}
  \bibinfo{year}{2001}\natexlab{}.
\newblock \showarticletitle{Emotion inferences from vocal expression correlate
  across languages and cultures}.
\newblock \bibinfo{journal}{\emph{Journal of Cross-Cultural Psychology}}
  \bibinfo{volume}{32}, \bibinfo{number}{1} (\bibinfo{date}{January}
  \bibinfo{year}{2001}), \bibinfo{pages}{76--92}.
\newblock


\bibitem[\protect\citeauthoryear{Scherer, Stratou, Gratch, Boberg, Mahmoud,
  Rizzo, and Morency}{Scherer et~al\mbox{.}}{2013}]%
        {SchererEtAl13_ABD}
\bibfield{author}{\bibinfo{person}{Stefan Scherer}, \bibinfo{person}{Giota
  Stratou}, \bibinfo{person}{Jonathan Gratch}, \bibinfo{person}{Jill Boberg},
  \bibinfo{person}{Marwa Mahmoud}, \bibinfo{person}{Albert~(Skip) Rizzo}, {and}
  \bibinfo{person}{Louis-Philippe Morency}.} \bibinfo{year}{2013}\natexlab{}.
\newblock \showarticletitle{Automatic Behavior Descriptors for Psychological
  Disorder Analysis}. In \bibinfo{booktitle}{\emph{Proc.\ 10th IEEE
  International Conference and Workshops on Automatic Face \& Gesture
  Recognition (FG)}}. \bibinfo{publisher}{IEEE}, \bibinfo{address}{Shanghai,
  P.\,R.\ China}.
\newblock
\newblock
\shownote{8 pages.}


\bibitem[\protect\citeauthoryear{Scherer, Stratou, Lucas, Mahmoud, Boberg,
  Gratch, Rizzo, and Morency}{Scherer et~al\mbox{.}}{2014}]%
        {scherer2014automatic}
\bibfield{author}{\bibinfo{person}{Stefan Scherer}, \bibinfo{person}{Giota
  Stratou}, \bibinfo{person}{Gale Lucas}, \bibinfo{person}{Marwa Mahmoud},
  \bibinfo{person}{Jill Boberg}, \bibinfo{person}{Jonathan Gratch},
  \bibinfo{person}{Albert~(Skip) Rizzo}, {and} \bibinfo{person}{Louis-Philippe
  Morency}.} \bibinfo{year}{2014}\natexlab{}.
\newblock \showarticletitle{Automatic audiovisual behavior descriptors for
  psychological disorder analysis}.
\newblock \bibinfo{journal}{\emph{Image and Vision Computing}}
  \bibinfo{volume}{32}, \bibinfo{number}{10} (\bibinfo{date}{October}
  \bibinfo{year}{2014}), \bibinfo{pages}{648--658}.
\newblock


\bibitem[\protect\citeauthoryear{Schmitt, Ringeval, and Schuller}{Schmitt
  et~al\mbox{.}}{2016}]%
        {Schmitt16-ATB}
\bibfield{author}{\bibinfo{person}{Maximilian Schmitt}, \bibinfo{person}{Fabien
  Ringeval}, {and} \bibinfo{person}{Bj\"orn Schuller}.}
  \bibinfo{year}{2016}\natexlab{}.
\newblock \showarticletitle{At the border of acoustics and linguistics:
  Bag-of-Audio-Words for the recognition of emotions in speech}. In
  \bibinfo{booktitle}{\emph{Proc.\ of INTERSPEECH 2016, 17th Annual Conference
  of the International Speech Communication Association}}.
  \bibinfo{publisher}{ISCA}, \bibinfo{address}{San Francisco, CA, USA},
  \bibinfo{pages}{495--499}.
\newblock


\bibitem[\protect\citeauthoryear{Schmitt and Schuller}{Schmitt and
  Schuller}{2017}]%
        {Schmitt17-OIT}
\bibfield{author}{\bibinfo{person}{Maximilian Schmitt} {and}
  \bibinfo{person}{Bj\"orn Schuller}.} \bibinfo{year}{2017}\natexlab{}.
\newblock \showarticletitle{{openXBOW -- Introducing the Passau Open-Source
  Crossmodal Bag-of-Words Toolkit}}.
\newblock \bibinfo{journal}{\emph{{Journal of Machine Learning Research}}}
  \bibinfo{volume}{18}, \bibinfo{number}{96} (\bibinfo{year}{2017}),
  \bibinfo{pages}{1--5}.
\newblock


\bibitem[\protect\citeauthoryear{Schuller, Steidl, Batliner, Marschik,
  Baumeister, Dong, Hantke, Pokorny, Rathner, Bartl-Pokorny, Einspieler, Zhang,
  Baird, Amiriparian, Qian, Ren, Schmitt, Tzirakis, and Zafeiriou}{Schuller
  et~al\mbox{.}}{2018}]%
        {schuller2018interspeech}
\bibfield{author}{\bibinfo{person}{Bj{\"o}rn Schuller}, \bibinfo{person}{Stefan
  Steidl}, \bibinfo{person}{Anton Batliner}, \bibinfo{person}{Peter~B.
  Marschik}, \bibinfo{person}{Harald Baumeister}, \bibinfo{person}{Fengquan
  Dong}, \bibinfo{person}{Simone Hantke}, \bibinfo{person}{Florian~B. Pokorny},
  \bibinfo{person}{Eva-Maria Rathner}, \bibinfo{person}{Katrin~D.
  Bartl-Pokorny}, \bibinfo{person}{Christa Einspieler}, \bibinfo{person}{Dajie
  Zhang}, \bibinfo{person}{Alice Baird}, \bibinfo{person}{Shahin Amiriparian},
  \bibinfo{person}{Kun Qian}, \bibinfo{person}{Zhao Ren},
  \bibinfo{person}{Maximilian Schmitt}, \bibinfo{person}{Panagiotis Tzirakis},
  {and} \bibinfo{person}{Stefanos Zafeiriou}.} \bibinfo{year}{2018}\natexlab{}.
\newblock \showarticletitle{{The INTERSPEECH 2018 Computational Paralinguistics
  Challenge:Atypical \& Self-Assessed Affect, Crying \& Heart Beats }}. In
  \bibinfo{booktitle}{\emph{Proc.\ of INTERSPEECH 2018, 19th Annual Conference
  of the International Speech Communication Association}}.
  \bibinfo{publisher}{ISCA}, \bibinfo{address}{Hyderabad, India},
  \bibinfo{pages}{122--126}.
\newblock


\bibitem[\protect\citeauthoryear{Schuller, Valstar, Eyben, Cowie, and
  Pantic}{Schuller et~al\mbox{.}}{2012}]%
        {Schuller12-A2T}
\bibfield{author}{\bibinfo{person}{Bj\"orn Schuller}, \bibinfo{person}{Michel
  Valstar}, \bibinfo{person}{Florian Eyben}, \bibinfo{person}{Roddy Cowie},
  {and} \bibinfo{person}{Maja Pantic}.} \bibinfo{year}{2012}\natexlab{}.
\newblock \showarticletitle{{AVEC 2012 -- The continuous Audio/Visual Emotion
  Challenge}}. In \bibinfo{booktitle}{\emph{{Proc.\ 14th ACM International
  Conference on Multimodal Interaction (ICMI)}}}. \bibinfo{publisher}{ACM},
  \bibinfo{address}{Santa Monica, CA, USA}, \bibinfo{pages}{449--456}.
\newblock


\bibitem[\protect\citeauthoryear{Schuller, Valstar, Eyben, McKeown, Cowie, and
  Pantic}{Schuller et~al\mbox{.}}{2011}]%
        {Schuller11-A2T}
\bibfield{author}{\bibinfo{person}{Bj\"orn Schuller}, \bibinfo{person}{Michel
  Valstar}, \bibinfo{person}{Florian Eyben}, \bibinfo{person}{Gary McKeown},
  \bibinfo{person}{Roddy Cowie}, {and} \bibinfo{person}{Maja Pantic}.}
  \bibinfo{year}{2011}\natexlab{}.
\newblock \showarticletitle{{AVEC 2011 -- The First International Audio/Visual
  Emotion Challenge}}. In \bibinfo{booktitle}{\emph{{Proc.\ 4th Biannual
  International Conference on Affective Computing and Intelligent Interaction
  (ACII)}}}, Vol.~\bibinfo{volume}{II}. \bibinfo{publisher}{Springer},
  \bibinfo{address}{Memphis, TN, USA}, \bibinfo{pages}{415--424}.
\newblock


\bibitem[\protect\citeauthoryear{Schwarz and Clore}{Schwarz and Clore}{1983}]%
        {Schwarz83-MMA}
\bibfield{author}{\bibinfo{person}{Norbert Schwarz} {and}
  \bibinfo{person}{Gerard~L. Clore}.} \bibinfo{year}{1983}\natexlab{}.
\newblock \showarticletitle{Mood, misattribution, and judgements of well-being:
  Informative and directive functions of affective states}.
\newblock \bibinfo{journal}{\emph{Journal of Personality and Social
  Psychology}} \bibinfo{volume}{45}, \bibinfo{number}{3}
  (\bibinfo{date}{September} \bibinfo{year}{1983}), \bibinfo{pages}{512--523}.
\newblock


\bibitem[\protect\citeauthoryear{Schwerdtfeger}{Schwerdtfeger}{2004}]%
        {schwerdtfeger2004predicting}
\bibfield{author}{\bibinfo{person}{Andreas Schwerdtfeger}.}
  \bibinfo{year}{2004}\natexlab{}.
\newblock \showarticletitle{Predicting autonomic reactivity to public speaking:
  don't get fixed on self-report data!}
\newblock \bibinfo{journal}{\emph{International Journal of Psychophysiology}}
  \bibinfo{volume}{52}, \bibinfo{number}{3} (\bibinfo{year}{2004}),
  \bibinfo{pages}{217--224}.
\newblock


\bibitem[\protect\citeauthoryear{Schwerdtfeger and Rathner}{Schwerdtfeger and
  Rathner}{2016}]%
        {schwerdtfeger2016ecological}
\bibfield{author}{\bibinfo{person}{Andreas~R Schwerdtfeger} {and}
  \bibinfo{person}{Eva-Maria Rathner}.} \bibinfo{year}{2016}\natexlab{}.
\newblock \showarticletitle{The ecological validity of the autonomic-subjective
  response dissociation in repressive coping}.
\newblock \bibinfo{journal}{\emph{Anxiety, Stress, \& Coping}}
  \bibinfo{volume}{29}, \bibinfo{number}{3} (\bibinfo{year}{2016}),
  \bibinfo{pages}{241--258}.
\newblock


\bibitem[\protect\citeauthoryear{Shan, Gong, and Mcowan}{Shan
  et~al\mbox{.}}{2009}]%
        {shanetal2009}
\bibfield{author}{\bibinfo{person}{Caifeng Shan}, \bibinfo{person}{Shaogang
  Gong}, {and} \bibinfo{person}{Peter~W Mcowan}.}
  \bibinfo{year}{2009}\natexlab{}.
\newblock \showarticletitle{Facial expression recognition based on Local Binary
  Patterns: A comprehensive study}.
\newblock \bibinfo{journal}{\emph{Image and Vision Computing}}
  \bibinfo{volume}{27}, \bibinfo{number}{6} (\bibinfo{year}{2009}),
  \bibinfo{pages}{803--816}.
\newblock


\bibitem[\protect\citeauthoryear{Shapiro and MacInnis}{Shapiro and
  MacInnis}{2002}]%
        {shapiro02}
\bibfield{author}{\bibinfo{person}{Stewart Shapiro} {and}
  \bibinfo{person}{Deborah~J. MacInnis}.} \bibinfo{year}{2002}\natexlab{}.
\newblock \showarticletitle{Understanding program-induced mood effects:
  Decoupling arousal from valence}.
\newblock \bibinfo{journal}{\emph{Journal of Advertising}}
  \bibinfo{volume}{31}, \bibinfo{number}{4} (\bibinfo{date}{May}
  \bibinfo{year}{2002}), \bibinfo{pages}{15--26}.
\newblock


\bibitem[\protect\citeauthoryear{Simonyan and Zisserman}{Simonyan and
  Zisserman}{2014}]%
        {simonyan2014very}
\bibfield{author}{\bibinfo{person}{Karen Simonyan} {and}
  \bibinfo{person}{Andrew Zisserman}.} \bibinfo{year}{2014}\natexlab{}.
\newblock \bibinfo{title}{Very deep convolutional networks for large-scale
  image recognition}.
\newblock \bibinfo{howpublished}{\url{https://arxiv.org/abs/1409.1556}}.
\newblock
\newblock
\shownote{14 pages.}


\bibitem[\protect\citeauthoryear{Stappen, Cummins, Messner, Baumeister,
  Dineley, and Schuller}{Stappen et~al\mbox{.}}{2019}]%
        {Stappen19-CMU}
\bibfield{author}{\bibinfo{person}{Lukas Stappen}, \bibinfo{person}{Nicholas
  Cummins}, \bibinfo{person}{Eva Messner}, \bibinfo{person}{Harald Baumeister},
  \bibinfo{person}{Judith Dineley}, {and} \bibinfo{person}{Bj{\"o}rn
  Schuller}.} \bibinfo{year}{2019}\natexlab{}.
\newblock \showarticletitle{Context Modelling Using Hierarchical Attention
  Networks for Sentiment and Self-assessed Emotion Detection in Spoken
  Narratives}. In \bibinfo{booktitle}{\emph{Proc.\ 44th IEEE International
  Conference on Acoustics, Speech and Signal Processing (ICASSP)}}.
  \bibinfo{publisher}{IEEE}, \bibinfo{address}{Brighton, United Kingdom},
  \bibinfo{pages}{6680--6684}.
\newblock


\bibitem[\protect\citeauthoryear{Trigeorgis, Ringeval, Brueckner, Marchi,
  Nicolaou, Schuller, and Zafeiriou}{Trigeorgis et~al\mbox{.}}{2016}]%
        {Trigeorgis16-AFE}
\bibfield{author}{\bibinfo{person}{George Trigeorgis}, \bibinfo{person}{Fabien
  Ringeval}, \bibinfo{person}{Raymond Brueckner}, \bibinfo{person}{Erik
  Marchi}, \bibinfo{person}{Mihalis~A. Nicolaou}, \bibinfo{person}{Bj{\"o}rn
  Schuller}, {and} \bibinfo{person}{Stefanos Zafeiriou}.}
  \bibinfo{year}{2016}\natexlab{}.
\newblock \showarticletitle{{Adieu features? End-to-end speech emotion
  recognition using a deep Convolutional Recurrent Network}}. In
  \bibinfo{booktitle}{\emph{Proc.\ 41st IEEE International Conference on
  Acoustics, Speech and Signal Processing (ICASSP)}}.
  \bibinfo{publisher}{IEEE}, \bibinfo{address}{Shanghai, P.\,R.\ China},
  \bibinfo{pages}{5200--5204}.
\newblock


\bibitem[\protect\citeauthoryear{Valstar, Gratch, Schuller, Ringeval, Cowie,
  and Pantic}{Valstar et~al\mbox{.}}{2016a}]%
        {Valstar16-SFA}
\bibfield{author}{\bibinfo{person}{Michel Valstar}, \bibinfo{person}{Jonathan
  Gratch}, \bibinfo{person}{Bj\"orn Schuller}, \bibinfo{person}{Fabien
  Ringeval}, \bibinfo{person}{Roddy Cowie}, {and} \bibinfo{person}{Maja
  Pantic}.} \bibinfo{year}{2016}\natexlab{a}.
\newblock \showarticletitle{{Summary for AVEC 2016: Depression, mood, and
  emotion recognition workshop and challenge}}. In
  \bibinfo{booktitle}{\emph{{Proc.\ 24th ACM International Conference on
  Multimedia (ACM MM)}}}. \bibinfo{publisher}{ACM},
  \bibinfo{address}{Amsterdam, The Netherlands}, \bibinfo{pages}{1483--1484}.
\newblock


\bibitem[\protect\citeauthoryear{Valstar, Gratch, Schuller, Ringeval, Lalanne,
  {Torres Torres}, Scherer, Stratou, Cowie, and Pantic}{Valstar
  et~al\mbox{.}}{2016b}]%
        {Valstar16-A2D}
\bibfield{author}{\bibinfo{person}{Michel Valstar}, \bibinfo{person}{Jonathan
  Gratch}, \bibinfo{person}{Bj\"orn Schuller}, \bibinfo{person}{Fabien
  Ringeval}, \bibinfo{person}{Denis Lalanne}, \bibinfo{person}{Mercedes {Torres
  Torres}}, \bibinfo{person}{Stefan Scherer}, \bibinfo{person}{Giota Stratou},
  \bibinfo{person}{Roddy Cowie}, {and} \bibinfo{person}{Maja Pantic}.}
  \bibinfo{year}{2016}\natexlab{b}.
\newblock \showarticletitle{{AVEC 2016 -- Depression, mood, and emotion
  recognition workshop and challenge}}. In \bibinfo{booktitle}{\emph{{Proc.\
  6th International Workshop on Audio/Visual Emotion Challenge (AVEC)}}}.
  \bibinfo{publisher}{ACM}, \bibinfo{address}{Amsterdam, The Netherlands},
  \bibinfo{pages}{3--10}.
\newblock


\bibitem[\protect\citeauthoryear{Valstar, Schuller, Krajewski, Cowie, and
  Pantic}{Valstar et~al\mbox{.}}{2013}]%
        {Valstar13-A2T}
\bibfield{author}{\bibinfo{person}{Michel Valstar}, \bibinfo{person}{Bj\"orn
  Schuller}, \bibinfo{person}{Jarek Krajewski}, \bibinfo{person}{Roddy Cowie},
  {and} \bibinfo{person}{Maja Pantic}.} \bibinfo{year}{2013}\natexlab{}.
\newblock \showarticletitle{{Workshop summary for the 3rd international
  Audio/Visual Emotion Challenge and workshop}}. In
  \bibinfo{booktitle}{\emph{Proc.\ 21st ACM International Conference on
  Multimedia (ACM MM)}}. \bibinfo{publisher}{ACM}, \bibinfo{address}{Barcelona,
  Spain}, \bibinfo{pages}{1085--1086}.
\newblock


\bibitem[\protect\citeauthoryear{Valstar, Schuller, Krajewski, Cowie, and
  Pantic}{Valstar et~al\mbox{.}}{2014}]%
        {Valstar14-A2T}
\bibfield{author}{\bibinfo{person}{Michel Valstar}, \bibinfo{person}{Bj\"orn
  Schuller}, \bibinfo{person}{Jarek Krajewski}, \bibinfo{person}{Roddy Cowie},
  {and} \bibinfo{person}{Maja Pantic}.} \bibinfo{year}{2014}\natexlab{}.
\newblock \showarticletitle{{AVEC 2014: The 4th international Audio/Visual
  Emotion Challenge and workshop}}. In \bibinfo{booktitle}{\emph{Proc.\ 22nd
  ACM International Conference on Multimedia (ACM MM)}}.
  \bibinfo{publisher}{ACM}, \bibinfo{address}{Orlando, FL, USA},
  \bibinfo{pages}{1243--1244}.
\newblock


\bibitem[\protect\citeauthoryear{Wataraka~Gamage, Dang, Sethu, Epps, and
  Ambikairajah}{Wataraka~Gamage et~al\mbox{.}}{2018}]%
        {Wataraka18-SBC}
\bibfield{author}{\bibinfo{person}{Kalani Wataraka~Gamage},
  \bibinfo{person}{Ting Dang}, \bibinfo{person}{Vidhyasaharan Sethu},
  \bibinfo{person}{Julien Epps}, {and} \bibinfo{person}{Eliathamby
  Ambikairajah}.} \bibinfo{year}{2018}\natexlab{}.
\newblock \showarticletitle{Speech-based Continuous Emotion Prediction by
  Learning Perception Responses Related to Salient Events: A Study Based on
  Vocal Affect Bursts and Cross-Cultural Affect in AVEC 2018}. In
  \bibinfo{booktitle}{\emph{{Proc.\ 8th International Workshop on Audio/Visual
  Emotion Challenge, AVEC'18}}}. \bibinfo{publisher}{ACM},
  \bibinfo{address}{Seoul, South Korea}, \bibinfo{pages}{47--55}.
\newblock


\bibitem[\protect\citeauthoryear{Weninger, Ringeval, Marchi, and
  Schuller}{Weninger et~al\mbox{.}}{2016}]%
        {Weninger16-DTR}
\bibfield{author}{\bibinfo{person}{Felix Weninger}, \bibinfo{person}{Fabien
  Ringeval}, \bibinfo{person}{Erik Marchi}, {and} \bibinfo{person}{Bj\"orn
  Schuller}.} \bibinfo{year}{2016}\natexlab{}.
\newblock \showarticletitle{{Discriminatively trained recurrent neural networks
  for continuous dimensional emotion recognition from audio}}. In
  \bibinfo{booktitle}{\emph{{Proc.\ 25th International Joint Conference on
  Artificial Intelligence (IJCAI)}}}. \bibinfo{publisher}{IJCAI/AAAI},
  \bibinfo{address}{New York City, NY, USA}, \bibinfo{pages}{2196--2202}.
\newblock


\bibitem[\protect\citeauthoryear{Williamson, Quatieri, Helfer, Horwitz, Yu, and
  Mehta}{Williamson et~al\mbox{.}}{2013}]%
        {williamsonetal13_vbd}
\bibfield{author}{\bibinfo{person}{James~R. Williamson},
  \bibinfo{person}{Thomas~F. Quatieri}, \bibinfo{person}{Brian~S. Helfer},
  \bibinfo{person}{Rachelle Horwitz}, \bibinfo{person}{Bea Yu}, {and}
  \bibinfo{person}{Daryush~D. Mehta}.} \bibinfo{year}{2013}\natexlab{}.
\newblock \showarticletitle{Vocal Biomarkers of Depression Based on Motor
  Incoordination}. In \bibinfo{booktitle}{\emph{Proc.\ 3rd ACM International
  Workshop on Audio/Visual Emotion Challenge}}
  \emph{(\bibinfo{series}{AVEC'13})}. \bibinfo{publisher}{ACM},
  \bibinfo{address}{Barcelona, Spain}, \bibinfo{pages}{41--48}.
\newblock


\bibitem[\protect\citeauthoryear{Yannakakis, Cowie, and Busso}{Yannakakis
  et~al\mbox{.}}{2017}]%
        {Yannakakis17-TON}
\bibfield{author}{\bibinfo{person}{Georgios~N. Yannakakis},
  \bibinfo{person}{Roddy Cowie}, {and} \bibinfo{person}{Carlos Busso}.}
  \bibinfo{year}{2017}\natexlab{}.
\newblock \showarticletitle{The ordinal nature of emotions}. In
  \bibinfo{booktitle}{\emph{{Proc.\ 7th Biannual Conference on Affective
  Computing and Intelligent Interaction (ACII)}}}. \bibinfo{publisher}{IEEE},
  \bibinfo{address}{San Antonio, TX}, \bibinfo{pages}{248--255}.
\newblock


\bibitem[\protect\citeauthoryear{Zhang, Provost, and Essl}{Zhang
  et~al\mbox{.}}{2017b}]%
        {Zhang17-CCA}
\bibfield{author}{\bibinfo{person}{Biqiao Zhang}, \bibinfo{person}{Emily~Mower
  Provost}, {and} \bibinfo{person}{Georg Essl}.}
  \bibinfo{year}{2017}\natexlab{b}.
\newblock \showarticletitle{Cross-corpus acoustic emotion recognition with
  multi-task learning: Seeking common ground while preserving differences}.
\newblock \bibinfo{journal}{\emph{IEEE Transactions on Affective Computing}}
  \bibinfo{volume}{10}, \bibinfo{number}{1} (\bibinfo{date}{March}
  \bibinfo{year}{2017}), \bibinfo{pages}{85--99}.
\newblock


\bibitem[\protect\citeauthoryear{Zhang, Cummins, and Schuller}{Zhang
  et~al\mbox{.}}{2017a}]%
        {Zhang17-EDE}
\bibfield{author}{\bibinfo{person}{Zixing Zhang}, \bibinfo{person}{Nicholas
  Cummins}, {and} \bibinfo{person}{Bj\"orn Schuller}.}
  \bibinfo{year}{2017}\natexlab{a}.
\newblock \showarticletitle{Advanced data exploitation in speech analysis -- An
  overview}.
\newblock \bibinfo{journal}{\emph{IEEE Signal Processing Magazine}}
  \bibinfo{volume}{34}, \bibinfo{number}{4} (\bibinfo{date}{July}
  \bibinfo{year}{2017}), \bibinfo{pages}{107--129}.
\newblock


\bibitem[\protect\citeauthoryear{Zhao, Li, Chen, and Jin}{Zhao
  et~al\mbox{.}}{2018}]%
        {Zhao18-MMD}
\bibfield{author}{\bibinfo{person}{Jinming Zhao}, \bibinfo{person}{Ruichen Li},
  \bibinfo{person}{Shizhe Chen}, {and} \bibinfo{person}{Qin Jin}.}
  \bibinfo{year}{2018}\natexlab{}.
\newblock \showarticletitle{Multi-modal Multi-cultural Dimensional Continues
  Emotion Recognition in Dyadic Interactions}. In
  \bibinfo{booktitle}{\emph{{Proc.\ 8th International Workshop on Audio/Visual
  Emotion Challenge, AVEC'18}}}. \bibinfo{publisher}{ACM},
  \bibinfo{address}{Seoul, South Korea}, \bibinfo{pages}{65--72}.
\newblock


\end{thebibliography}

\end{document}